# Band Gap Modulation of $SrTiO_3$ upon $CO_2$ Adsorption

*Kostiantyn V. Sopiha[1], Oleksandr I. Malyi[2], Clas Persson[2,3], Ping Wu[1,\*]*

1 – Engineering Product Development, Singapore University of Technology and Design, 8 Somapah Road, 487372 Singapore, Singapore

2 – Centre for Materials Science and Nanotechnology, University of Oslo, P. O. Box 1048 Blindern, NO-0316 Oslo, Norway

3 – Department of Physics, University of Oslo, P. O. Box 1048 Blindern, NO-0316 Oslo, Norway

E-mail: wuping@sutd.edu.sg (W.P)

**Abstract**

$CO_2$ chemisorption on $SrTiO_3(001)$ surfaces is studied using *ab initio* calculations in order to establish new chemical sensing mechanisms. We find that $CO_2$ adsorption opens the material band gap, however, while the adsorption on the $TiO_2$-terminated surface neutralizes surface states at the valence band (VB) maximum, $CO_2$ on the SrO-terminated surfaces suppresses the conduction band (CB) minimum. For the $TiO_2$-terminated surface, the effect is explained by the passivation of dangling bonds, whereas for the SrO-terminated surface, the suppression is caused by the surface relaxation. Modulation of the VB states implies a more direct change in charge distribution, and thus the induced change in band gap is more prominent at the $TiO_2$ termination. Further, we show that both $CO_2$ adsorption energy and surface band gap are strongly dependent on $CO_2$ coverage, suggesting that the observed effect can be utilized for sensing application in a wide range of $CO_2$ concentrations.

**Introduction**

$CO_2$ monitoring is of significant interest for various applications including breath analysis,[1] blood diagnostic,[2] gas monitoring,[3] and climate control.[4] Current $CO_2$ sensing technologies rely mainly on optical and electrochemical sensing mechanisms.[5, 6] Despite recent advances, manufacturing of optical sensors remains too expensive for a wide range of applications.[5] At the same time, the usage of electrochemical $CO_2$ sensors is limited to high-temperature environments.[6] As such, the development of cost-effective chemiresistive $CO_2$ detectors is believed to be the ultimate goal in the field.[7] From a materials perspective, it requires chemically active material having strong and selective interaction with $CO_2$. However, an outstanding chemical stability of $CO_2$ makes it difficult to ionize upon adsorption and, therefore, the number of systems capable of $CO_2$ sensing is very limited. Although a few traditional sensing materials (such as $ZnO$,[8] $SnO_2$,[9] $WO_3$,[10] $La_2O_3$[11]) have been utilized for $CO_2$ detection, reported sensitivity is too low for practical usage.[7] At the same time, various perovskites[12-17] and perovskite-based composites[18] demonstrate a profound $CO_2$ chemiresistive effect. Owing to their stability, electronic properties, and low production cost, this class of materials has received significant attention in sensing technologies. During the last few years, molecular adsorption on different perovskite surfaces was investigated by both experimental[19-24] and first-principles methods.[21, 25-28] The results provided a basic understanding of the surface chemistry of perovskites, revealing rather strong $CO_2$ adsorption even on stoichiometric surfaces. Interestingly, it was also found that the electronic structures of some perovskite surfaces differ from that in bulk, which is mainly attributed to surface states entering the band gap.[29-32] This property can lead to the increase in the surface conductivity compared to that in bulk,[33-35] making the overall electronic properties of nanoscaled perovskites sensitive to chemical reactions occurring at their surfaces. In particular, Duan et al. have shown that adsorption of atomic hydrogen on SrTiO$_3$(001) surfaces suppresses the surface states.[36] However, since only atomic adsorption was considered, the obtained results cannot be directly linked to the non-ionizing molecular adsorption. In one of the recent works, Fei et al. demonstrated that adsorption of H$_2$O and CO$_2$ affects the band gap energy of K$_{1-Y}$Na$_Y$Ta$_{1-X}$Nb$_X$O$_3$(001) due to the suppression of the surface states.[37, 38] These studies illustrate a strong potential of perovskite surfaces for sensing application. Despite that, a complete understanding of the effect as well as its origin has not been established yet.

By virtue of its outstanding chemical stability, SrTiO$_3$ has been known for its ability to sense O$_2$,[39, 40] H$_2$,[41] and various hydrocarbons[42] without being affected by the harsh operation conditions. Recently, Dai et al. took one step forward to utilize a two-dimensional electron gas at LaAlO$_3$/SrTiO$_3$ interface for fast and selective detection of various oxidizing and reducing gases at room temperature.[43] However, despite presence of previous experimental reports on both CO$_2$ uptake[19-21] and band gap reduction[33, 44] on bare SrTiO$_3$(001) surfaces, a corresponding CO$_2$ sensing mechanism is still to be developed. Motivated by this, in this work, we perform a detailed *ab initio* study of CO$_2$ adsorption on the most stable SrTiO$_3$ surfaces.[29, 45] The obtained results are expected to guide a further development of CO$_2$ sensing technologies.

**Methods**

All presented results are obtained using Vienna ab-initio simulation package (VASP) with exchange-correlation interaction described by the Perdew-Burke-Ernzerhof (PBE) functional.[46] Projector augmented wave (PAW) pseudopotentials[47, 48] are used to model the effect of core electrons. Strontium 4s$^2$4p$^6$5s$^2$, titanium 3s$^2$3p$^6$3d$^2$4s$^2$, oxygen 2s$^2$2p$^4$, and carbon 2s$^2$2p$^2$ electrons are treated explicitly. The cutoff energy for plane-wave basis sets is varied from 300 to 500 eV for screening and final relaxations, respectively, unless otherwise specified. $\Gamma$-centered Monkhorst-Pack[49] grids of different sizes (see Table S1) are used for the Brillouin-zone integrations. For PBE+U calculations, Hubbard U correction[50] of 4.2 eV is used for $d$ orbitals of Ti atoms. Obtained results are analyzed with Vesta[51] and pymatgen.[52]

The analysis of $CO_2$ adsorption chemistry is carried out using 2×2 supercell slabs of cubic (Pm$\overline{3}$m) $SrTiO_3$ containing 13 atomic layers and 15 Å of vacuum region. Here, the $SrTiO_3$ slab is constructed based on the computed lattice constants of 3.94 and 3.97 Å for PBE and PBE+U calculations, respectively, which are quite close to the experimental value of 3.90 Å.[40] This model provides reasonable accuracy for calculations of both surface band gap and $CO_2$ adsorption energies on $SrTiO_3$(001) surfaces (see Figs. S1 and S2). To investigate $CO_2$ coverage effects, lateral sizes of the supercell are varied as given in Table S1. Both SrO- and $TiO_2$-terminated $SrTiO_3$(001) surfaces are considered. To avoid dipole-dipole interaction[53] between the SrO- and $TiO_2$-terminated sides of the slab, we use non-stoichiometric SrO-SrO and $TiO_2$-$TiO_2$ slab systems. This approach is widely accepted for systems having surface dipole moments, and for the $SrTiO_3$ slabs in particular.[26, 29, 45, 54] Moreover, symmetrical $CO_2$ adsorption on opposite sides of the slab is adopted for all calculations. To exclude transition between cubic (Pm$\overline{3}$m) and tetragonal (I4/mcm) phases of $SrTiO_3$, only four top layers for both sides of the slab are free to relax upon the adsorption while other five bulk-like atomic layers are kept fixed. Similar fixation of the deep bulk-like layers is commonly adopted for studying molecular adsorption on $SrTiO_3$(001) surfaces, but the number of relaxed surface layers is generally less.[21, 25] As it can be seen in Figs. S1 and S2, our approach provides more accurate description of the adsorption and band gap energies.

To find the stable adsorption configurations, we analyze a uniform grid of $CO_2$ adsorption positions on both SrO- and $TiO_2$-terminated $SrTiO_3$(001) surfaces considering horizontal, vertical, and $45^o$-tilted allocations of the molecule. For each configuration, the minimum distance from the unrelaxed outermost atomic layer of the $SrTiO_3$ slab to the closest O atom of the $CO_2$ molecule is adjusted to 2 Å. This method provides about 130 nonequivalent adsorption configurations on each termination. Moreover, to account for the existence of more complex adsorption conformations, we also study an adsorption of neutral carbonate $CO_3$-like complex near a surface oxygen vacancy. Here, the periodic interaction between the surface defects is minimized by increasing a lateral size of the modeled supercell to 3×3 unit cells (in contrast to 2×2 supercell used elsewhere). The initial adsorption positions are generated via random rotation and in-plane displacement of the $CO_3$ complex, located 2.5 Å above and parallel to the $SrTiO_3$(001) surfaces containing one oxygen vacancy. In total, we consider more than 20 nonequivalent $CO_3$-like adsorption configurations for each termination. Next, the initial configurations are run through the series of ionic relaxations until the atomic forces are less than 0.01 eV/Å. To understand thermodynamics of $CO_2$ interaction with the $SrTiO_3$(001) surfaces, adsorption energy is calculated as $\mathrm{E_{Ads}} = \left(\mathrm{E}(\mathrm{slab} + n \cdot \mathrm{CO_2}) - \mathrm{E}(\mathrm{slab}) - n \cdot \mathrm{E}(\mathrm{CO_2})\right)/n$, where $\mathrm{E}(\mathrm{slab} + n \cdot \mathrm{CO_2})$, $\mathrm{E}(\mathrm{slab})$, and $\mathrm{E}(\mathrm{CO_2})$ are total energies of the slab containing $n$ adsorbed $CO_2$ molecules, bare relaxed slab after $CO_2$ desorption, and free $CO_2$ molecule, respectively. It should be pointed out that the reference states used here correspond to the $SrTiO_3$ slabs after $CO_2$ removal. This is mainly because the deformed surface may not recover after a $CO_2$ desorption completely due to a partial phase transition between the cubic and tetragonal crystal symmetries of $SrTiO_3$.

Bader charge transfer to X ion (X = Sr, Ti, O, and C) is calculated as $Q(X) = Q(X) - Q_{ref}(X)$, where $Q(X)$ and $Q_{ref}(X)$ are the Bader charges[55-58] of X in the studied and in the reference systems, respectively. Clean $SrTiO_3(001)$ slabs and free $CO_2$ molecule are taken as the reference systems. The Bader charge of X ion is taken as positive (negative) if the atom loses (gains) electrons, e.g. the computed Bader charges of Sr, Ti, and O ions in bulk $SrTiO_3$ are 1.59e, 2.04e, and -1.21e, respectively. These values are consistent with previous theoretical studies[30, 31] and reflect presence of both more (Sr-O) and less (Ti-O) ionic chemical bonds in $SrTiO_3$.[59, 60] To quantify surface relaxation, we employ structure parameters corresponding to changes in the interplanar distances between the first and the second atomic planes ($d_{12}$), the second and the third atomic planes ($d_{23}$), as well as an average outwards displacement of O with respect to the metal ions in the first ($S_1$), and the second ($S_2$) atomic layers (see Fig. S3). Positions of the atomic planes are determined as an average z coordinate of the metal ions in the considered atomic layer. To compare the electronic structures of the $SrTiO_3(001)$ surfaces having different work functions,[30, 61] alignments of electronic energy levels are estimated based on the average electrostatic potential at cores of Ti atoms in the bulk-like regions.

### Results and discussion

First, we perform a detailed study of clean $SrTiO_3(001)$ surfaces to understand their electronic properties. Numerous experimental studies have revealed that the (001) surface of as-grown $SrTiO_3$ appears as a random mixture of both $TiO_2$ and SrO terminations,[33, 44] and their ratio can be controlled by varying material synthesis conditions[62] as well as using simple post-synthesis treatments.[63, 64] Indeed, previous *ab initio* studies confirmed that the SrO- and $TiO_2$-terminated surfaces have close energetics.[29, 45] Similar to previous reports, we find that both surfaces experience significant relaxations as summarized in Table S2. Moreover, we observe that the TiO$_2$-terminated surface obtained from an ionic relaxation of as-cleaved cubic SrTiO$_3$ is metastable with respect to the surface reconstruction having a clear geometry pattern as given in Fig. S4. It should be noted that the computed energy difference per 2×2 surface of the reconstructed and as-relaxed (not reconstructed) TiO$_2$-terminated surfaces is 0.11 eV. At the same time, no surface reconstruction for SrO-terminated surfaces is observed. Therefore, the reconstructed TiO$_2$-terminated surface is taken as a reference state and addressed as a "clean surface" henceforth.

Despite similar energetics, the electronic properties of the SrO- and TiO$_2$-terminated surfaces are very different (see Fig. 1). In particular, the band gap energy of TiO$_2$-terminated slab is 0.99 eV, which is significantly smaller than 1.80 eV for bulk SrTiO$_3$. This difference is explained by 2$p$ orbitals of O[26, 29, 31, 65] at the outermost TiO$_2$ layer (1-TiO$_2$) emerging energetically above the VBM of bulk SrTiO$_3$ (see Figs. 1a,c). The surface localization of these states is also illustrated by the partial charge density distribution in Fig. 1b. The observed modifications of electronic structures are caused by the formation of dangling bonds at the as-cleaved (not relaxed) TiO$_2$-terminated surface partially passivated by the surface relaxation; this is displayed in Fig. S5. For the SrO-terminated surface, the reduction in band gap energy is also observed, but it is less notable as compared to that for the TiO$_2$-terminated surface. Here, the outermost SrO layer (1-SrO) determines neither the VBM nor the conduction band minimum (CBM) of the slab system (see Fig. 1e). The reduction comes from the emergence of vacant Ti $3d$ orbitals at energies near CBM[31, 65] for the subsurface TiO$_2$ layer (2-TiO$_2$), see Fig. 1g. This is caused by the relaxation of Ti-O bonds between Ti atoms in 2-TiO$_2$ and O atoms in the adjacent layers, which is due to the surface relaxation (see Fig. S5). For instance, the band gap energy of as-cleaved surface is only 0.02 eV smaller as compared to the bulk value, while for the relaxed slab it decreases to 1.55 eV. The localization of surface states is also confirmed by the analysis of charge densities at VBM and CBM as shown in Fig. 1f.

Since the formation of the surface states at $SrTiO_3(001)$ is governed by the Ti-O bonding,[45, 54, 66] accurate description of Ti *3d* orbitals is critical. However, classical PBE functional is known to describe localized states inaccurately.[67] This limitation is reflected, in part, by highly underestimated band gap energy of 1.80 eV obtained from the PBE calculations as compared to the experimental value of 3.25 eV.[68] To make sure that the appearance of the surface states is not sensitive to the simulation method, here we also compute the electronic properties of the $SrTiO_3$ surfaces using PBE+U method with the Hubbard U correction applied to *d* orbitals of the Ti atoms. Interestingly, we notice that the observed surface reconstruction of the $TiO_2$-terminated surface (see Fig. S4) becomes unstable when the correction is applied, relaxing into the ideal 1×1 surface supercell as obtained by relaxation of as-cleaved $SrTiO_3(001)$ surface. Moreover, we also find that the U correction increases the band gap energy of bulk $SrTiO_3$ from 1.80 to 2.28 eV, while for the clean $TiO_2$-terminated surface, the corresponding increase in the band gap energy is from 0.99 to 1.53 eV. At the same time, the band gap energy of the SrO-terminated slab changes from 1.55 to 2.15 eV when U correction is applied. These results clearly demonstrate that the computed band gap energies are sensitive to U correction. Despite that, the computed differences in the band gap energies for the bulk and $SrTiO_3(001)$ surfaces are comparable; that is, 0.81 (0.25) and 0.75 (0.13) eV for PBE and PBE+U calculations of the $TiO_2$(SrO)-terminated slabs, respectively. Importantly, the similar reductions are also observed for a wide range of Hubbard U parameters as illustrated in Fig. S6, indicating that the formation of surface states can be qualitatively described by either the PBE or PBE+U approaches (see

Figs. 1d,h). Therefore, the following results refer to the PBE calculations, unless otherwise specified.

By following the screening procedure for $CO_2$ adsorption on the SrO-terminated $SrTiO_3(001)$, we obtain five stable adsorption configurations corresponding to the formation of $CO_3$-like complex (see Figs. 2a, S7). The lowest energy configuration obtained by the screening has a $C_s$ point group symmetry of $CO_3$-like complex and is 0.49 eV more stable as compared to the previously reported by Baniecki et al. with the $C_{2v}$ point group symmetry,[21, 25] which is also reproduced here (see Fig. S7c). Moreover, we also find two other configurations that are more stable as compared to that presented by Baniecki et al.[21, 25] (see Figs. S7a,b). It worth nothing that formation of $CO_3$-like complexes is common for oxide surfaces.[69-75] In the most stable conformation, the C atom is chemically bonded to the surface oxygen $O_S$ with the short bond of 1.33 Å, which is shorter than a typical length of single C-O bond (about 1.43 Å)[76] but noticeably larger than that computed for $CO_2$ molecule (1.18 Å). At the same time, each O atom of the $CO_2$ molecule forms two separate bonds with the nearest Sr atoms. The first bond length of 2.72 Å is comparable to the Sr-O bond length in bulk (2.79 Å), while the second bond of 2.54 Å is significantly shorter. The $CO_2$ molecule is also deformed upon adsorption. Specifically, the interaction stretches original C-O bond from 1.18 to 1.29 Å and reduces O-C-O angle from $180^o$ to $122^o$ indicating strong chemical bonding of adsorbed $CO_2$ molecule to the surface. Indeed, computed adsorption energy of -1.94 eV and the Bader charge transfer (see Methods) of 0.38e from the adsorbed molecule evince stable $CO_2$ chemisorption on this surface.

Unlike for the SrO-terminated surface, the screening protocol applied to the $TiO_2$-terminated surface results in only one stable chemisorptions position as illustrated in the Fig. 2b. This configuration is equivalent to the previously reported by Baniecki et al.[21, 25] and corresponds to the formation of a $CO_3$-like complex with the $C_{2v}$ point group symmetry, where the C and O atoms of the $CO_2$ molecule are strongly bonded to the surface oxygen $O_S$ and nearest Ti atoms, respectively. The formed Ti-O bond of 2.10 Å is slightly longer than that in bulk $SrTiO_3$ (1.97 Å), while the C-$O_S$ bond length is about 1.37 Å, i.e., similar to the corresponding bond for the SrO-termination. The $CO_2$ molecule is also deformed here. In particular, the original C-O bond in stretched from 1.18 to 1.27 Å, while O-C-O angle is reduced from 180 to 131$^\circ$. It is important to note that the adsorption energy of -1.24 eV and the Bader charge transfer of 0.1e to the $TiO_2$-terminated surface indicate weaker $CO_2$ bonding than that for the SrO-terminated surface.

To better understand the adsorption chemistry, we also investigate the Bader charge redistribution within the separated as-deformed slabs and $CO_2$ molecules with respect to the clean slabs and free $CO_2$ molecule. As shown in Fig. S8, the C (O) atoms of the as-deformed $CO_2$ molecules have significantly lower (higher) Bader charge than in the free molecule. It clearly indicates that the C-O bonds in the deformed molecules are much weaker as compared to a free $CO_2$ molecule. When attached to the surface, the C and O atoms form chemical bonds with the surface oxygen $O_S$ and cations (Sr or Ti atoms), respectively, compensating the Bader charge redistribution imposed by the $CO_2$ deformation. At the same time, charge distributions for $SrTiO_3$ slabs with and without the molecule are similar, with the major difference in total Bader charge transfers of 0.38 and 0.10e to the SrO- and $TiO_2$-terminated surfaces, respectively (see Figs. S8a,b).

The formed chemical bonding between $CO_2$ and $SrTiO_3$(001) surfaces has a significant impact on their electronic structures (see Fig. 3). In particular, the $CO_2$ adsorption on the SrO-terminated surface leads to a noticeable increase in the band gap energy from 1.55 to 1.82 eV. It is mainly attributed to the suppression of the surface states at the CBM level of the 2-$TiO_2$ layer. As one can also notice, the computed band gap energy of the slab exceeds that of the bulk $SrTiO_3$ (1.80 eV). This is nothing but a model limitation, which can be easily eliminated by using thicker $SrTiO_3$ slab system. However, it implies that the $CO_2$ adsorption is able to eliminate the surface states in this system completely. Moreover, the charge density redistribution induced by the adsorption results in significant modification of the surface geometry as summarized in Table S3. As one can see, it results in decreasing the structure parameters ($d_{12}$, $d_{23}$, $S_1$, and $S_2$), suggesting that the surface geometries become closer to those of as-cleaved unrelaxed $SrTiO_3$(001) slab. Since the reduction in the band gap energy is originated from the atomic relaxation, it also suggests that the observed suppression of surface states upon $CO_2$ adsorption is in large part originated from the surface relaxation. This effect is well illustrated in Fig. S5, demonstrating that the surface states can be partially suppressed by artificial fixing the structure parameters to the values obtained from the $CO_2$ adsorption analysis (Table S3). $CO_2$ adsorption on the $TiO_2$-terminated surface also leads to significant increase in the band gap energy from 0.99 to 1.50 eV (see Figs. 3e,g). Not only this change is more prominent as compared to that for the SrO-terminated surface, but it has also a different origin. The effect is mainly attributed to the passivation of dangling bonds, resulting in suppression of the corresponding surface states at the clean $TiO_2$-terminated surface. As one would expect, a major change is observed at the 1-$TiO_2$ layer while the bulk-like region is barely affected. The $CO_2$ adsorption also affects the structure parameters (see Table S3), but the influence of that relaxation on the surface band gap energy is negligible (see Fig. S5).

Similar to the clean slab, $CO_2$ adsorption on both $SrTiO_3$(001) surfaces was also studied by PBE+U approach using the lowest energy configurations obtained from the PBE calculations.

Computed adsorption energies are -1.84 and -1.22 eV for the SrO- and $TiO_2$-terminated surfaces, respectively. These values are close to those obtained without Hubbard U correction (-1.94 and -1.24 eV). In addition, the correction increases the band gap energy of the $CO_2$-containing system to 2.29 and 1.93 eV for the SrO- and $TiO_2$-terminated surfaces, respectively (see Figs. 3d,h). These values are 0.14 and 0.40 eV larger than the corresponding band gap energies of the clean slabs, but 0.01 eV larger and 0.35 eV smaller than that of bulk $SrTiO_3$ (2.28 eV), respectively. Furthermore, as it can be seen in Figs. S6b,c and S9, similar band gap openings upon $CO_2$ adsorption are also observed for other Hubbard U values as well as hybrid HSE[77] calculations. These results suggest that not only formation of the surface states, but also the effect of band gap opening upon $CO_2$ adsorption can be qualitatively described with either the PBE potential or the PBE+U approach and, therefore, only PBE computations presented henceforth.

To reveal how the band gap depends on $CO_2$ coverage, we model $CO_2$ adsorption on the $SrTiO_3(001)$ slabs with different supercell sizes corresponding to $\Theta = 0.111$ to 0.50 surface coverage range (see Table S1). Here, $\Theta$ is defined as the number of $CO_2$ molecules per unit cell of the surface. For the highest analyzed coverage of $\Theta = 0.5$, different coverage modes are analyzed as shown in Fig. S10. Studying the $TiO_2$-terminated surface, we find that throughout the analyzed coverage range the adsorption energy is strongly negative varying from -1.19 eV for $\Theta = 0.5$ to -1.57 eV for $\Theta = 0.125$. Despite a general trend of increasing with coverage (see Fig. 4a), the computed adsorption energy is not a monotonic function of the coverage, but rather of the lateral supercell size along O1-O2 of the adsorbed $CO_2$ molecule (see Figs. 4b, S11; Table S1). This dependence is the result of anisotropic surface stress imposed by $CO_2$ adsorption on the $TiO_2$-terminated surface. The band gap energy also increases with the coverage due to the suppression of the surface states upon $CO_2$ adsorption (see Figs. 4d,e). The largest energy gap of 1.85 eV at $\Theta = 0.5$ is 0.05 eV larger than the bulk value (1.80 eV), implying that the surface states at the $TiO_2$-terminated surface are fully passivated by the adsorption. The smallest energy gap of 1.19 eV at $\Theta = 0.125$ is 0.20 eV larger than that of the clean $TiO_2$-terminated slab (0.99 eV), suggesting that further reduction of the band gap energy at lower $CO_2$ coverage can be expected.

While studying the $CO_2$ adsorption energy on SrO-terminated $SrTiO_3(001)$, we find that the most stable adsorption configurations for high and low $CO_2$ coverage ranges are different. In particular, the most stable $CO_2$ configuration discussed above is only stable at $\Theta \geq 0.25$, but slowly relaxes to form another adsorption conformation with near $C_{3v}$ point group symmetry of $CO_3$-like complex at lower $CO_2$ coverage. It this adsorption configuration a surface oxygen atom $O_S$ is significantly displaced from its initial positions to form a $CO_3$-like complex adsorbed on the surface oxygen vacancy as illustrated in Fig. S7a. In fact, the formation of the surface oxygen vacancy in the small systems is probably the main reason why this configuration is metastable at high coverage; e.i. because of a periodic interaction between the surface defects. It is also interesting to note an appearance of filled in-gap defect states in $\sqrt{2} \times \sqrt{2}$ supercell resulting in band gap reduction to 1.43 eV as reflected in Figs. S12, 4d. It can also be speculated that this adsorption geometry and electronic structure provide more possibilities for a heterogeneous catalytic pathway and, therefore, this configuration is favorable for experimentally observed $CO_2$ reduction.[19, 78, 79] However, further and deeper research in this area is needed.

As can be seen in Fig. 4c, $CO_2$ adsorption energy on SrO-terminated $SrTiO_3(001)$ is a nearly linear function of coverage, varying from -1.46 eV for $\Theta = 0.5$ to -2.28 eV for $\Theta = 0.111$. These values are 0.3-0.7 eV lower than those for the $TiO_2$-terminated surface, suggesting stronger $CO_2$ adsorption on the SrO-terminated surface. Here, the band gap energy also increases with the coverage, reaching 1.84 eV at $\Theta = 0.5$. This value also exceeds the band gap energy of bulk $SrTiO_3$ (1.80 eV), ensuring complete elimination of the surface states upon $CO_2$ adsorption. Among the studied $CO_2$ coverages, the smallest band gap energy is 1.74 eV at $\Theta = 0.111$, which is 0.19 eV larger than that of the clean SrO-terminated slab. Therefore, it can be expected that lower $CO_2$ coverage will result in further decrease of the energy gap.

The observed band gap modulation represents a new approach for resistivity-based sensing applications; this is illustrated in Fig. 5. It is well known that electrical conductivity of semiconductors is determined by multiple factors, where the band gap energy is one of the most important parameters. Therefore, it can be speculated that the existing surface states at $SrTiO_3$(001) can explain the experimentally observed enhancement of electronic conductivity[33, 34, 44] and even contribute to the formation of two-dimensional electron gas at bare $SrTiO_3$(001).[80, 81] Here, we demonstrate that $CO_2$ adsorption leads to the suppression of the surface states, presumably decreasing the surface conductivity of $SrTiO_3$(001). According to our results, the effect is sensitive to the surface termination, and higher $CO_2$ response for the $TiO_2$-terminated surface is expected. Because of this, we suggest that better sensing performance can be achieved by controlling the surface termination. Based on experimental reports,[63, 64] this can be achieved by treating the crystal surface with a pH-controlled $NH_4F$-HF solution. Moreover, since the effect is localized to the surface layers, the response can be greatly enhanced for nanoscale systems. This simple strategy provides guidance to develop a new generation of chemical sensors which, unlike the common state-of-art gas sensors, do not require ionization of the adsorbed molecules and, therefore, can be used for sensing of chemically stable gas species.

**Conclusions**

Using *ab initio* calculations, we performed a detailed study of $CO_2$ adsorption on the $SrTiO_3(001)$ surfaces. We found that the adsorption results in the formation of highly stable $CO_3$-like complexes, where the lowest energy adsorption configuration is 0.49 eV more stable as compared to the reported by Baniecki et al.[21, 25] In the considered $\Theta = 0.111\text{-}0.50$ coverage range, the computed $CO_2$ adsorption energies vary from -1.57 (-2.28) to -1.19 (-1.46) eV for $TiO_2$-terminated (SrO-terminated) $SrTiO_3(001)$ surface. We also demonstrated that the adsorption leads to suppression of surface states at the clean $SrTiO_3(001)$ surfaces, which opens the material energy gap. In particular, for the $TiO_2$-terminated surface, the band gap energy is a monotonic function of the $CO_2$ coverage, and the gap energy changes from 0.99 to 1.85 eV for the bare and fully passivated surfaces, respectively. This effect is more prominent than that for the SrO-terminated $SrTiO_3(001)$, where the corresponding change in the energy gap is from 1.55 eV to 1.84 eV only. The developed understanding of the band gap modulation provides new opportunities to design resistivity-based $CO_2$ sensors.

**Acknowledgments**

The authors acknowledge the support from the SUTD-ZJU (ZJURP1200101), MOE Tier2 (T2 MOE1201-Singapore), and Research Council of Norway (contracts: 221469, 250346). Most this work was performed on the Abel cluster, owned by the University of Oslo and the Norwegian Metacenter for Computational Science (NOTUR), and operated by the Department for Research Computing at USIT, the University of Oslo IT-department. The authors also acknowledge PRACE for awarding access to resource MareNostrum based in Spain at BSC-CNS.

**Electronic Supplementary Information (ESI) available:** convergence of the computed (Fig. S1) band gap energy and (Fig. S2) $CO_2$ adsorption energy with model parameters of the SrTiO$_3$(001) slabs; (Fig. S3) illustration of the structure relaxation parameters used in this work; (Fig. S4) illustration of the surface reconstruction on TiO$_2$-terminated SrTiO$_3$(001) surface; (Fig. S5) formation and evolution of surface states at the SrTiO$_3$(001) surfaces; (Fig. S6) effect of Hubbard U parameter on the band gap energies of clean and CO$_2$-covered SrTiO$_3$(001) surfaces; (Fig. S7) metastable CO$_2$ adsorption configurations on the SrO-terminated surface; (Fig. S8) Bader charge transfer induced by CO$_2$ adsorption; (Fig. S9) projected densities of states of SrTiO$_3$(001) surfaces computed using hybrid HSE functional; (Fig. S10) different CO$_2$ coverage modes corresponding to $\Theta = 0.5$; (Fig. S11) illustration of "supercell size along O1-O2" term used in this manuscript; (Fig. S12) computed electronic structures of CO$_3$-like complex adsorbed on the surface oxygen vacancy in $\sqrt{2}\times\sqrt{2}$ SrO-terminated SrTiO$_3$(001) system; (Table S1) sizes of the supercells and corresponding Monkhorst-Pack grids used in the calculations; structure parameters for the (Table S2) clean and (Table S3) CO$_2$-containing SrTiO$_3$(001) slabs; (Table S4) summary of the computational results; stable CO$_2$ configurations in Crystallographic Information File (CIF) format.

**Figures, Tables, and Captions**

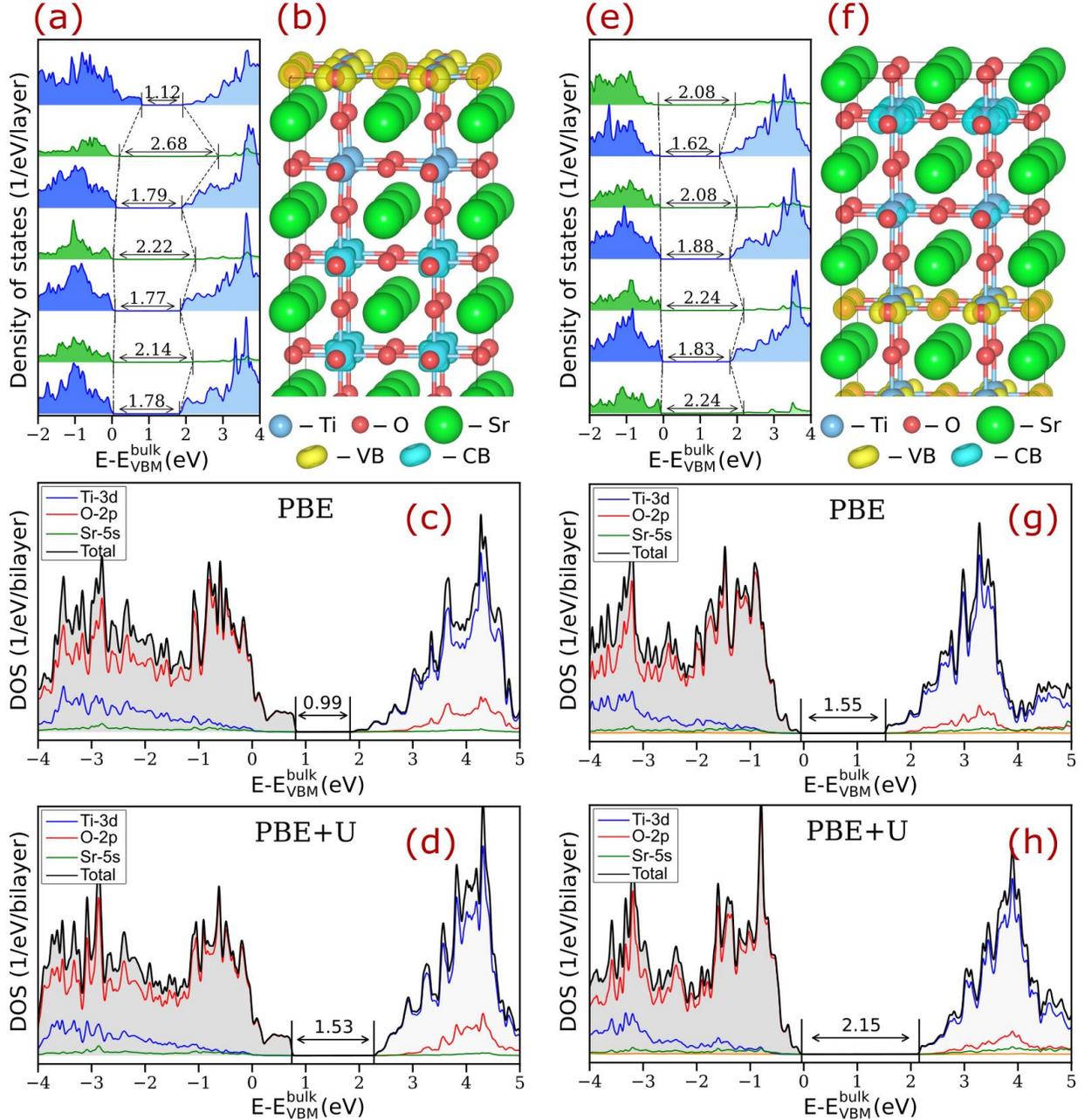

Figure 1. Computed electronic structures of clean (a-d) $TiO_2$- and (e-h) SrO-terminated $SrTiO_3$(001) surfaces. (a,e) Layer-resolved density of states (LDOS). Green and blue curves represent population densities for the SrO and $TiO_2$ layers, respectively. Numbers represent effective band gap energies of each atomic layer computed from LDOS neglecting the population densities below 0.1 1/eV/layer. (b,f) Partial charge density distributions at VB (yellow) and CB (blue) of the $SrTiO_3$ slabs. Projected density of states for two top layers

computed by PBE (c,g) and PBE+U (d,h) methods and band gap energies of the corresponding slabs. $E_{VBM}^{bulk}$ stands for VBM energy of bulk $SrTiO_3$.

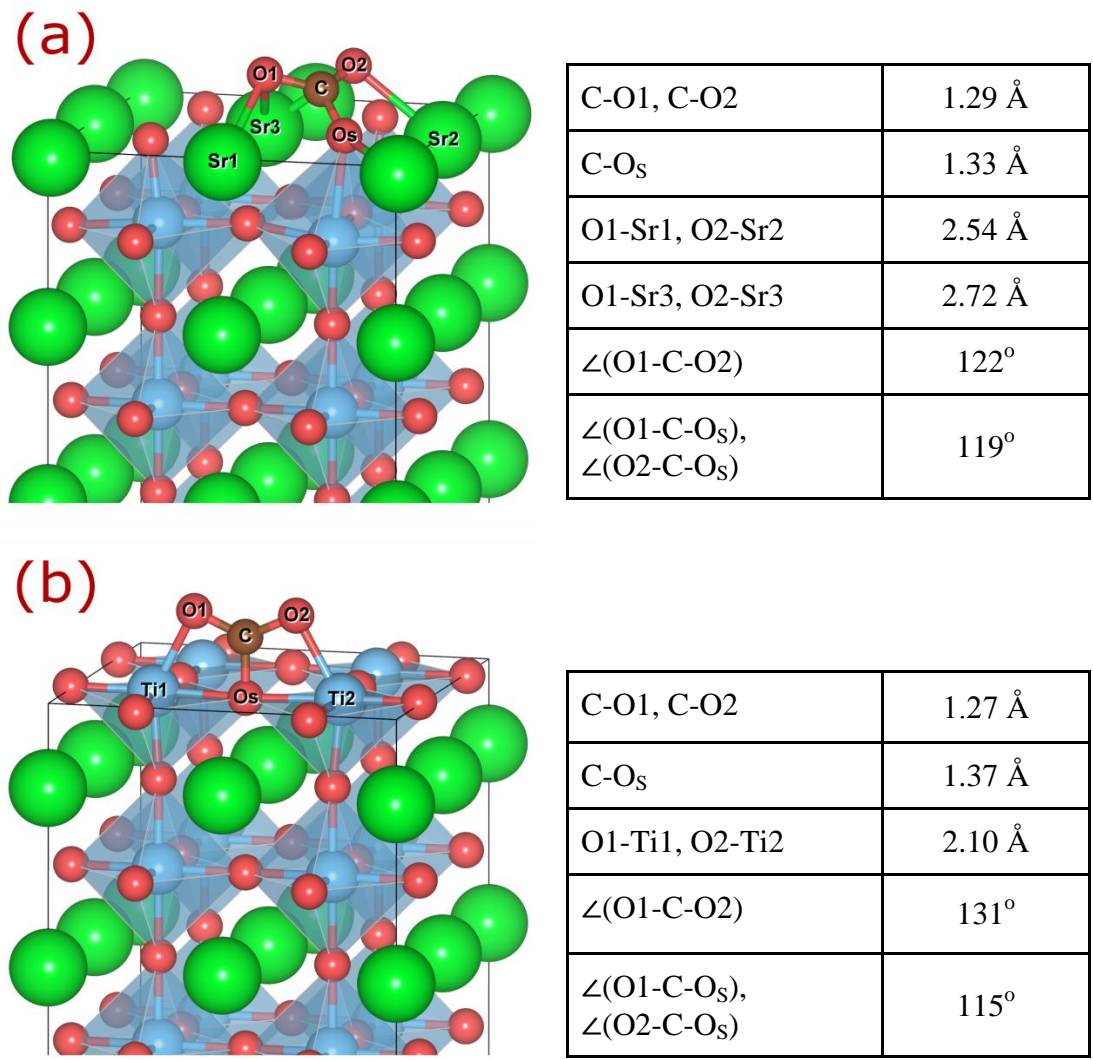

**(a)**

| C-O1, C-O2 | 1.29 Å |
|---|---|
| C-O$_S$ | 1.33 Å |
| O1-Sr1, O2-Sr2 | 2.54 Å |
| O1-Sr3, O2-Sr3 | 2.72 Å |
| $\angle$(O1-C-O2) | 122° |
| $\angle$(O1-C-O$_S$), $\angle$(O2-C-O$_S$) | 119° |

**(b)**

| C-O1, C-O2 | 1.27 Å |
|---|---|
| C-O$_S$ | 1.37 Å |
| O1-Ti1, O2-Ti2 | 2.10 Å |
| $\angle$(O1-C-O2) | 131° |
| $\angle$(O1-C-O$_S$), $\angle$(O2-C-O$_S$) | 115° |

Figure 2. Most stable $CO_2$ adsorption configurations on (a) SrO- and (b) TiO$_2$-terminated SrTiO$_3$(001) computed for 2×2 surface supercell.

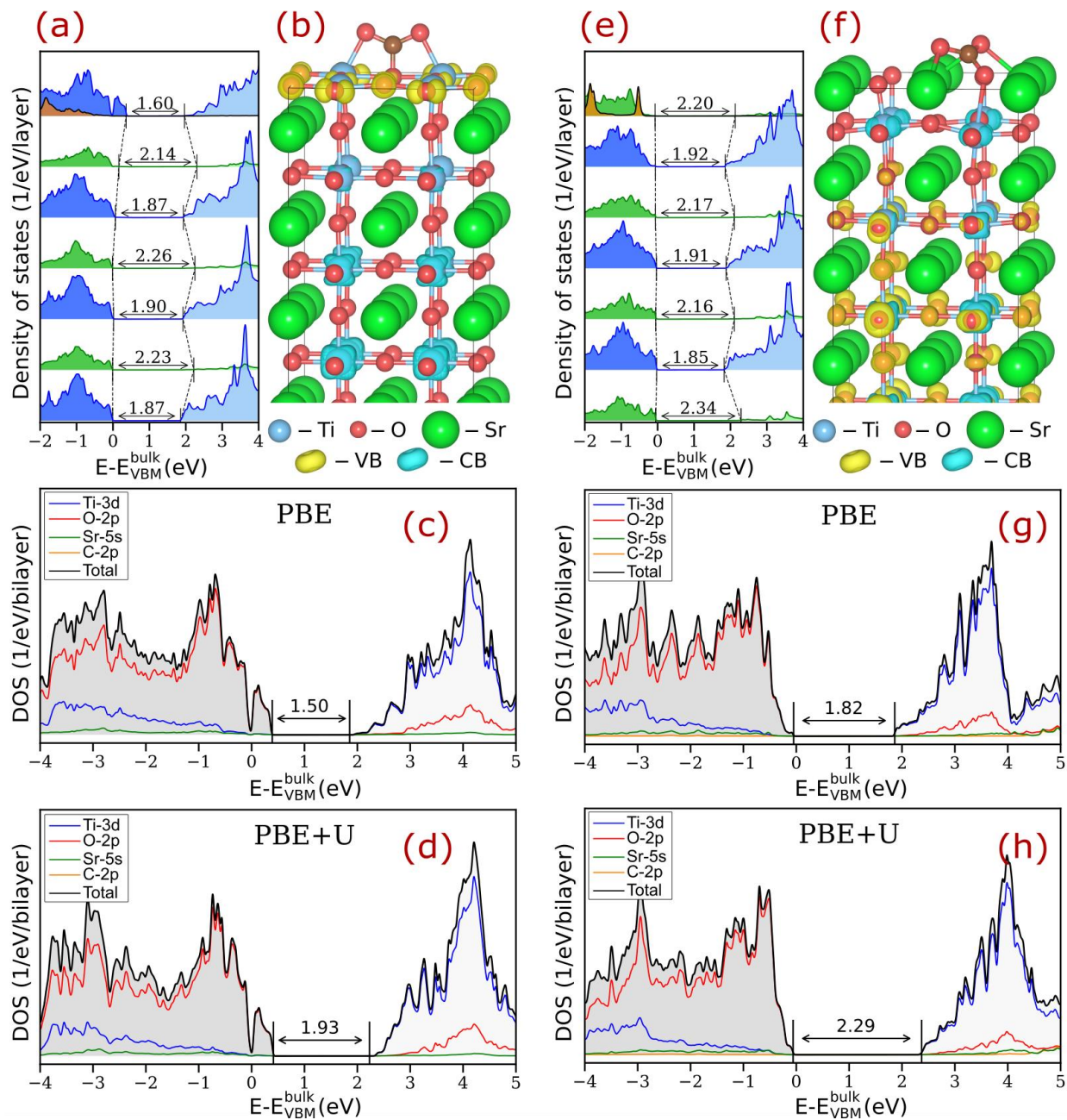

Figure 3. Computed electronic structures of (a-d) TiO$_2$- and (e-h) SrO-terminated SrTiO$_3$(001) surfaces containing one adsorbed CO$_2$ molecule per 2×2 surface supercell. (a,e) Layer-resolved density of states (LDOS). Orange, green, and blue curves represent population densities for the adsorbed CO$_2$ molecules, SrO, and TiO$_2$ layers, respectively. Numbers represent effective band gap energies of each atomic layer computed from LDOS neglecting the population densities below 0.1 1/eV/layer. (b,f) Partial charge density distributions at VB (yellow) and CB (blue) of the SrTiO$_3$ slabs. Projected density of states for two top layers containing adsorbed CO$_2$

molecule computed by PBE (c,g) and PBE+U (d,h) methods and band gap energies of the corresponding slabs. $E_{VBM}^{bulk}$ stands for VBM energy of bulk $SrTiO_3$.

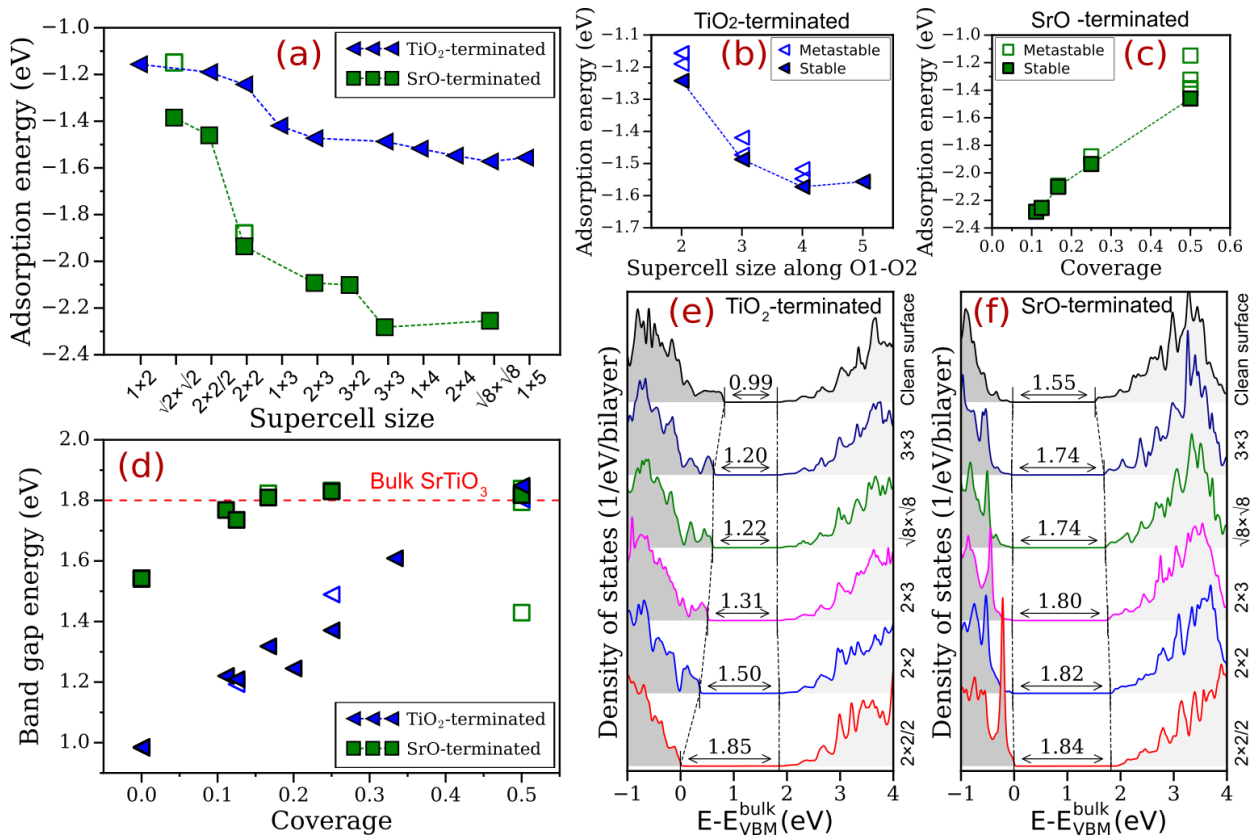

Figure 4. (a) Dependence of $CO_2$ adsorption energy on supercell size. Open markers represent metastable adsorption configuration. The systems are labeled by their lateral dimensions in units of lattice constant. Dependence of $CO_2$ adsorption energy on (b) lateral supercell size along O1-O2 direction (in units of the lattice constant, a = 3.94 Å) for the $TiO_2$-terminated surface and (c) $CO_2$ coverage for the SrO-terminated surface. (d) Coverage dependence of the band gap energy. Density of states for two top layers of $SrTiO_3$ slabs with different supercells of the (e) $TiO_2$-terminated and (f) SrO-terminated surfaces. Numbers indicate band gap energies of corresponding $SrTiO_3(001)$ slabs. $E_{VBM}^{bulk}$ stands for VBM energy of bulk $SrTiO_3$.

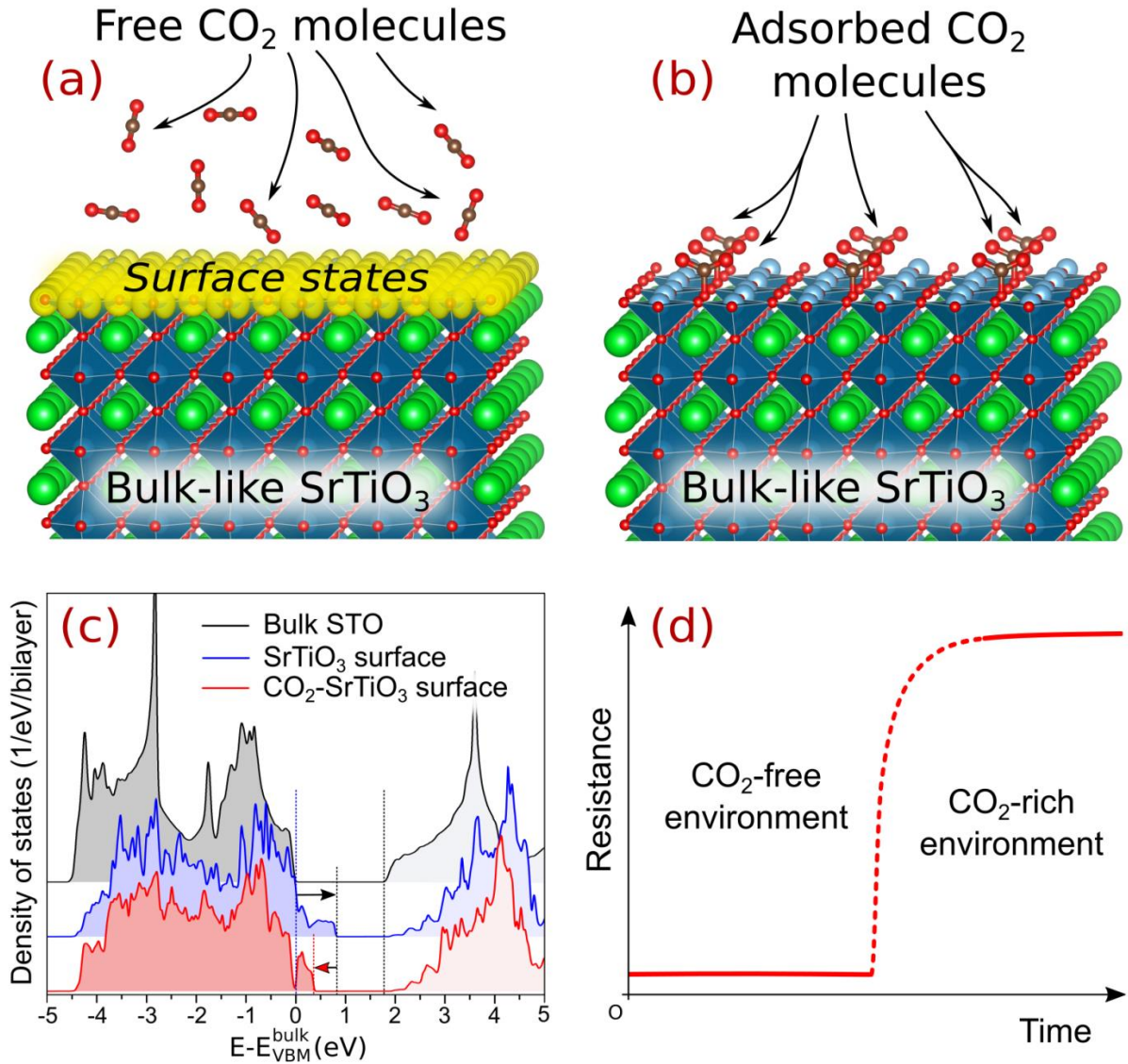

Figure 5. Schematic illustration of $CO_2$ sensing mechanism for $TiO_2$-terminated $SrTiO_3(001)$ surface. (a) Surface states at clean surface and (b) their suppression by $CO_2$ adsorption. (c) Corresponding modification of the electronic structure given for two top layers at the surface. (d) Schematic illustration of $CO_2$ sensing response utilizing band gap modulation effect.

**Supplementary information:**

**Band Gap Modulation of SrTiO$_3$ upon CO$_2$ Adsorption**

*Kostiantyn V. Sopiha[1], Oleksandr I. Malyi[2], Clas Persson[2,3], Ping Wu[1,\*]*

1 – Engineering Product Development, Singapore University of Technology and Design, 8 Somapah Road, 487372 Singapore, Singapore

2 – Centre for Materials Science and Nanotechnology, University of Oslo, P. O. Box 1048 Blindern, NO-0316 Oslo, Norway

3 – Department of Physics, University of Oslo, P. O. Box 1048 Blindern, NO-0316 Oslo, Norway

E-mail: wuping@sutd.edu.sg (W.P)

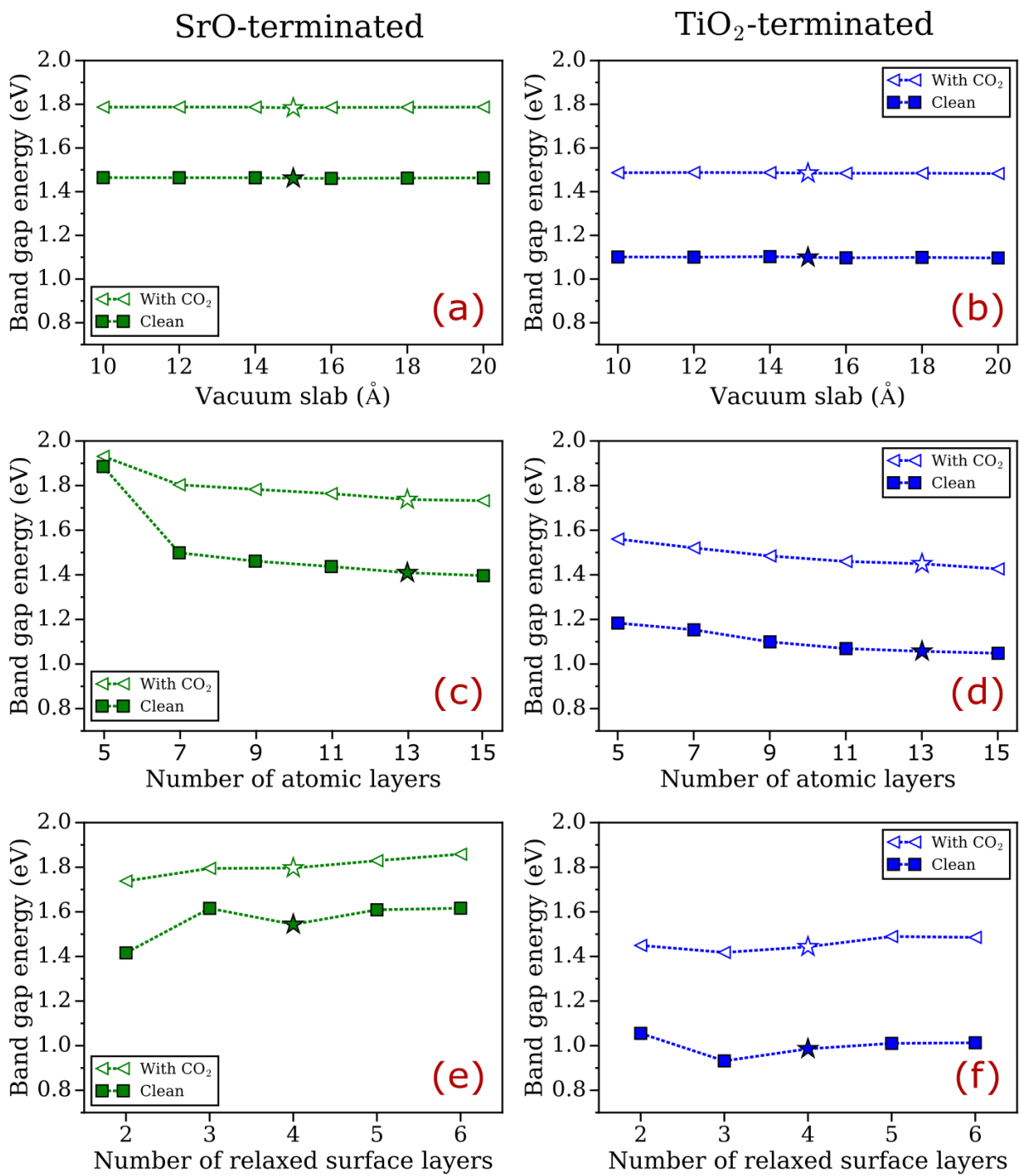

Figure S1. Convergence of the computed band gap energy with (a,b) thickness of vacuum slab, (c,d) number of atomic layers, and (e,f) number of relaxed surface layers for modeled (a,c,e) SrO-terminated and (b,d,f) TiO$_2$-terminated SrTiO$_3$(001) slabs. Star markers represent default values of X used in all other calculations. (a-d) are computed for SrTiO$_3$(001) slabs with two relaxed surface layers. Other model parameters are described in the methods.

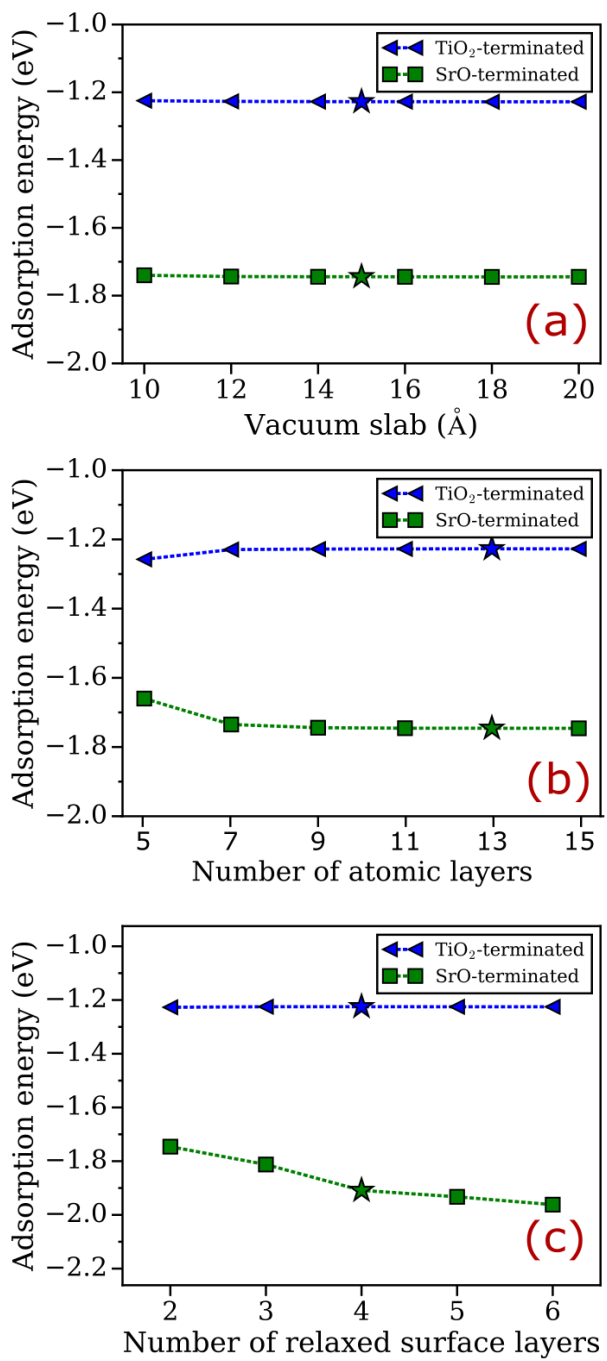

Figure S2. Convergence of the computed $CO_2$ adsorption energy with (a) thickness of vacuum slab, (b) number of atomic layers, and (c) number of relaxed surface layers for modeled $SrTiO_3(001)$ slab. Star markers represent default values of X used in all other calculations. (a,b) are computed for $SrTiO_3(001)$ slabs with two relaxed surface layers. Other model parameters are described in the methods.

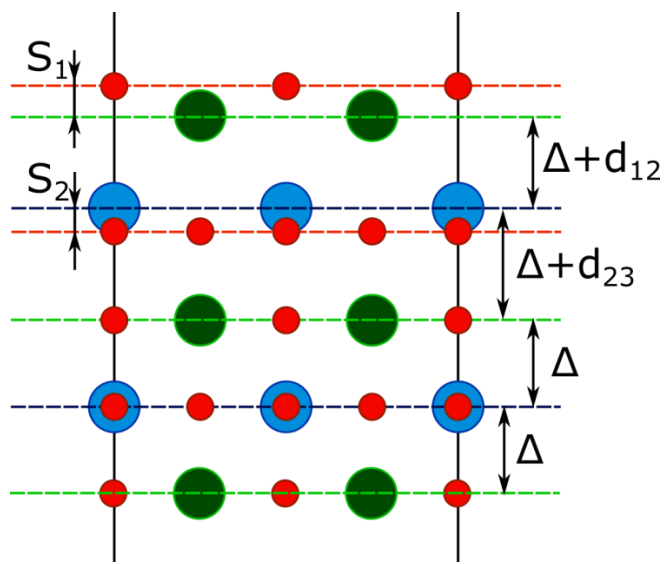

Figure S3. Illustration of the structure relaxation parameters used in this work. $\Delta$ represents the interplanar distance in bulk $SrTiO_3$ ($2\Delta = 3.94$ Å). Green, blue, and red circles represent Sr, Ti, and O ions, respectively.

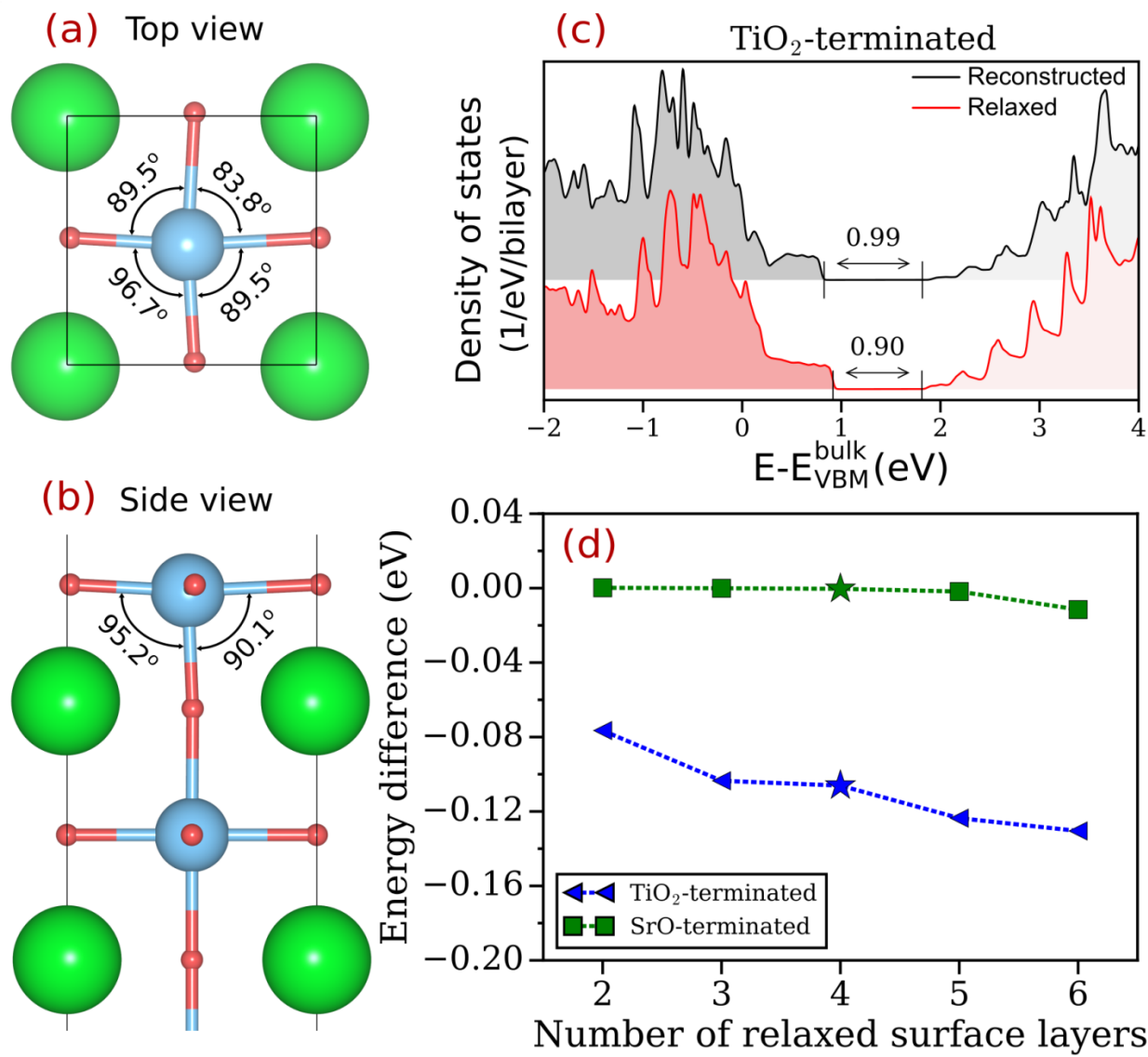

Figure S4. Illustration of the surface reconstruction on TiO$_2$-terminated SrTiO$_3$(001) surface. (a) top view and (b) side view of the reconstructed surface. (c) Comparison of electronic structures of relaxed (not reconstructed) and reconstructed surfaces. (d) Energy difference per 2×2 surface supercell between relaxed (not reconstructed) and reconstructed SrTiO$_3$(001) surface vs number of relaxed surface layers. Star markers represent default values of X used in all other calculations.

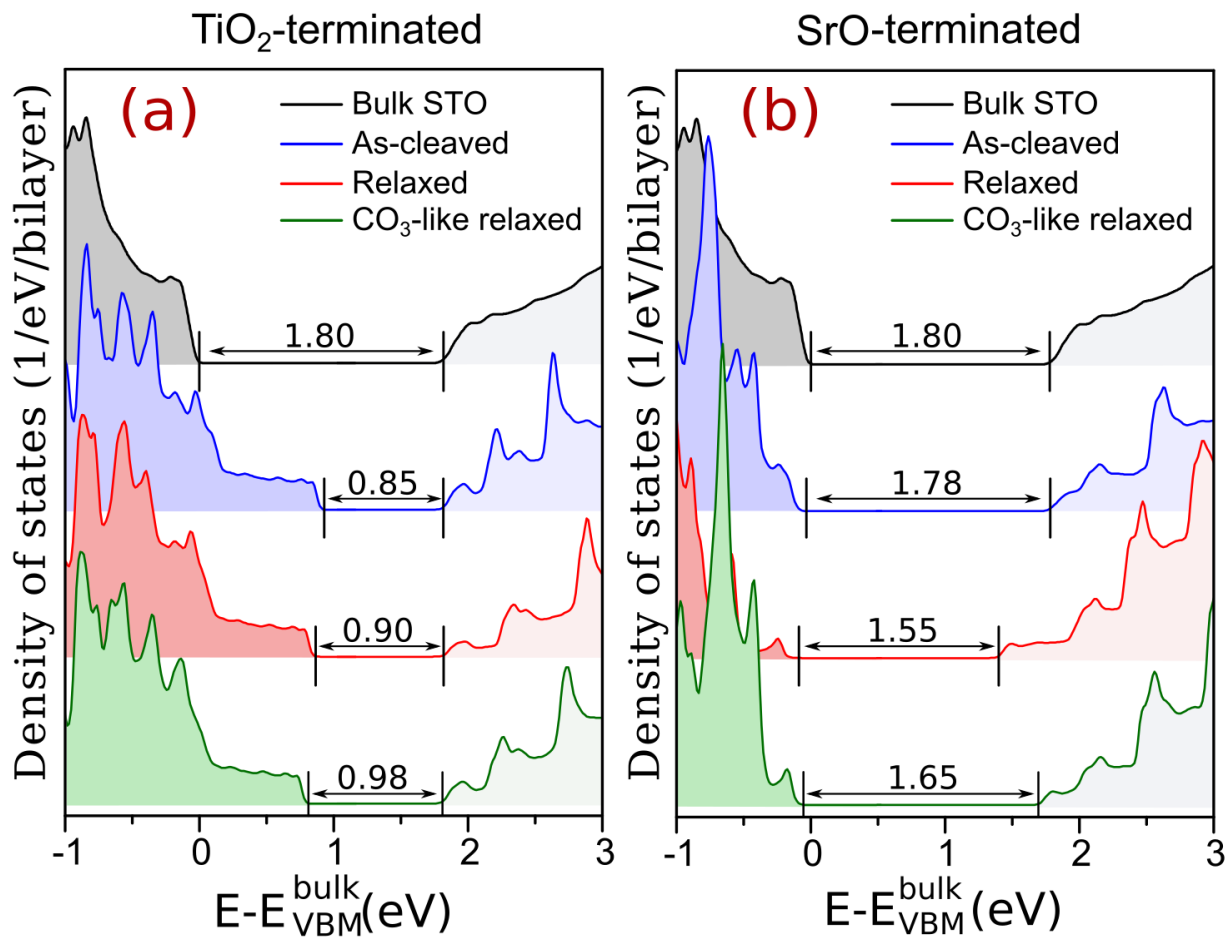

Figure S5. Formation and evolution of the surface states at (a) TiO$_2$-terminated and (b) SrO-terminated SrTiO$_3$(001). Black, red, blue, and green curves represent densities of states for bulk SrTiO$_3$, as-cleaved, relaxed (not reconstructed), and "CO$_3$-like relaxed" surfaces, respectively. Structure parameters of "CO$_3$-like relaxed" surface are identical to those of the SrTiO$_3$(001) containing one adsorbed CO$_2$ molecule per 2×2 surface supercell (see Table S3). Strokes and numbers represent valence band maximum (VBM), conduction band minimum (CBM), and band gap energies of SrTiO$_3$(001) slabs.

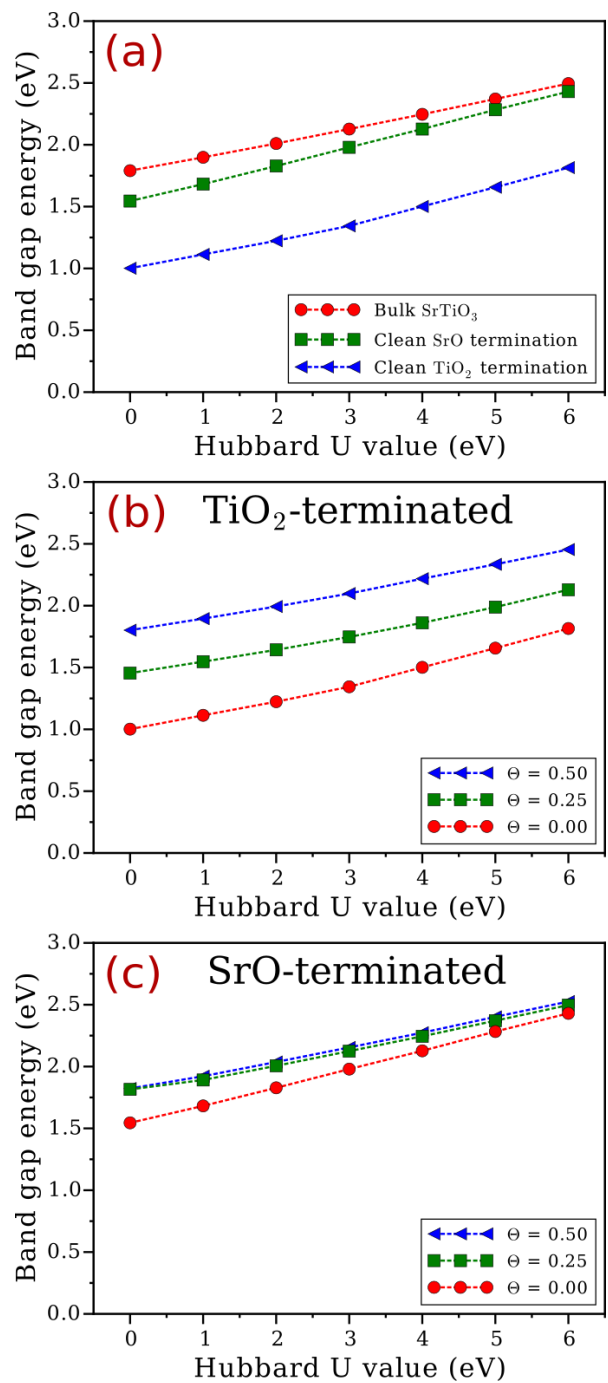

Figure S6. (a) Band gap energies of bulk SrTiO$_3$ and clean TiO$_2$- and SrO-terminated SrTiO$_3$(001) surfaces as functions of Hubbard U value. Band gap energies of (b) TiO$_2$- and (c) SrO-terminated SrTiO$_3$(001) surfaces as functions of Hubbard U value for slab systems containing zero ($\Theta = 0.00$), one ($\Theta = 0.25$), and two ($\Theta = 0.50$) adsorbed CO$_2$ molecules per 2×2 surface supercell. Here, cutoff energy of 400 eV for plane wave basis sets was used for all DFT+U calculations.

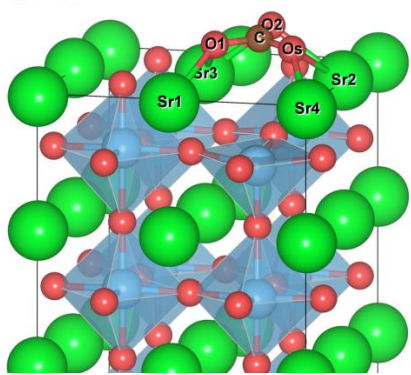

| C-O1, C-O2, C-O$_S$ | 1.30 Å |
|---|---|
| O1-Sr1, O$_S$-Sr4 | 2.44 Å |
| O1-Sr3, O$_S$-Sr2 | 2.65 Å |
| O2-Sr3, O2-Sr2 | 2.56 Å |
| ∠(O1-C-O2) | 122° |
| ∠(O1-C-O2), ∠(O2-C-O$_S$) | 119° |

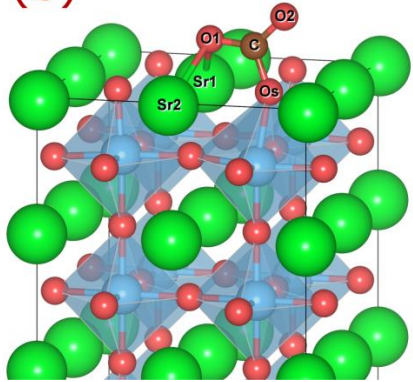

| C-O1 | 1.30 Å |
|---|---|
| C-O2 | 1.25 Å |
| C-O$_S$ | 1.38 Å |
| O1-Sr1, O1-Sr2 | 2.63 Å |
| ∠(O1-C-O2) | 129° |
| ∠(O1-C-O$_S$) | 115° |
| ∠(O2-C-O$_S$) | 117° |

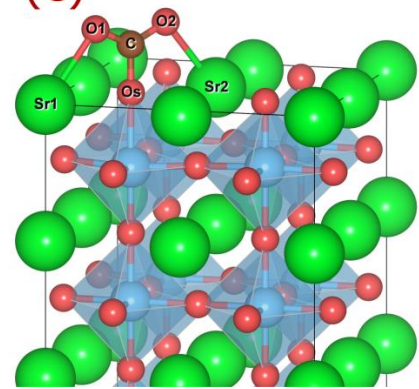

| C-O1, C-O2 | 1.27 Å |
|---|---|
| C-O$_S$ | 1.39 Å |
| O1-Sr1, O2-Sr2 | 2.60 Å |
| ∠(O1-C-O2) | 129° |
| ∠(O1-C-O$_S$), ∠(O2-C-O$_S$) | 115° |

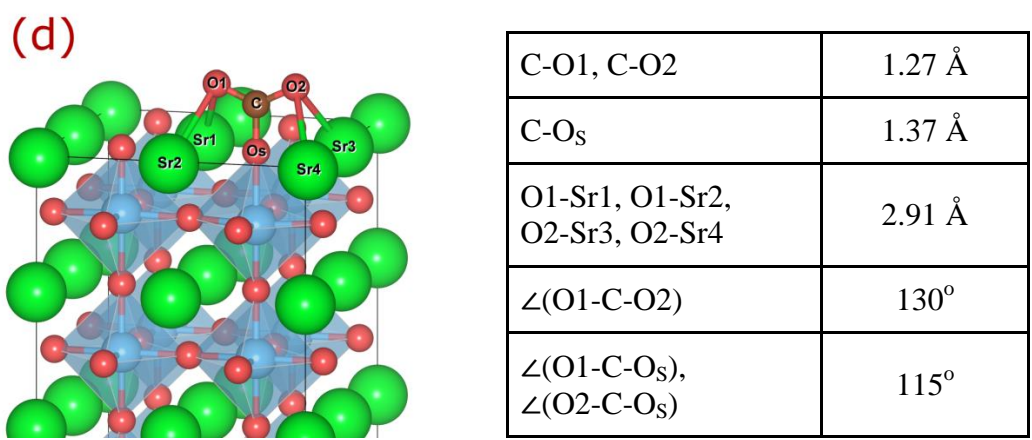

| | |
|---|---|
| C-O1, C-O2 | 1.27 Å |
| C-O$_S$ | 1.37 Å |
| O1-Sr1, O1-Sr2, O2-Sr3, O2-Sr4 | 2.91 Å |
| ∠(O1-C-O2) | 130° |
| ∠(O1-C-O$_S$), ∠(O2-C-O$_S$) | 115° |

Figure S7. Metastable $CO_2$ adsorption configurations on SrO-terminated $SrTiO_3$(001) surface. The computed adsorption energies for (a), (b), (c), and (d) are 0.06, 0.47, 0.49, and 0.54 eV higher as compared to the most stable adsorption configuration presented in Fig. 2a.

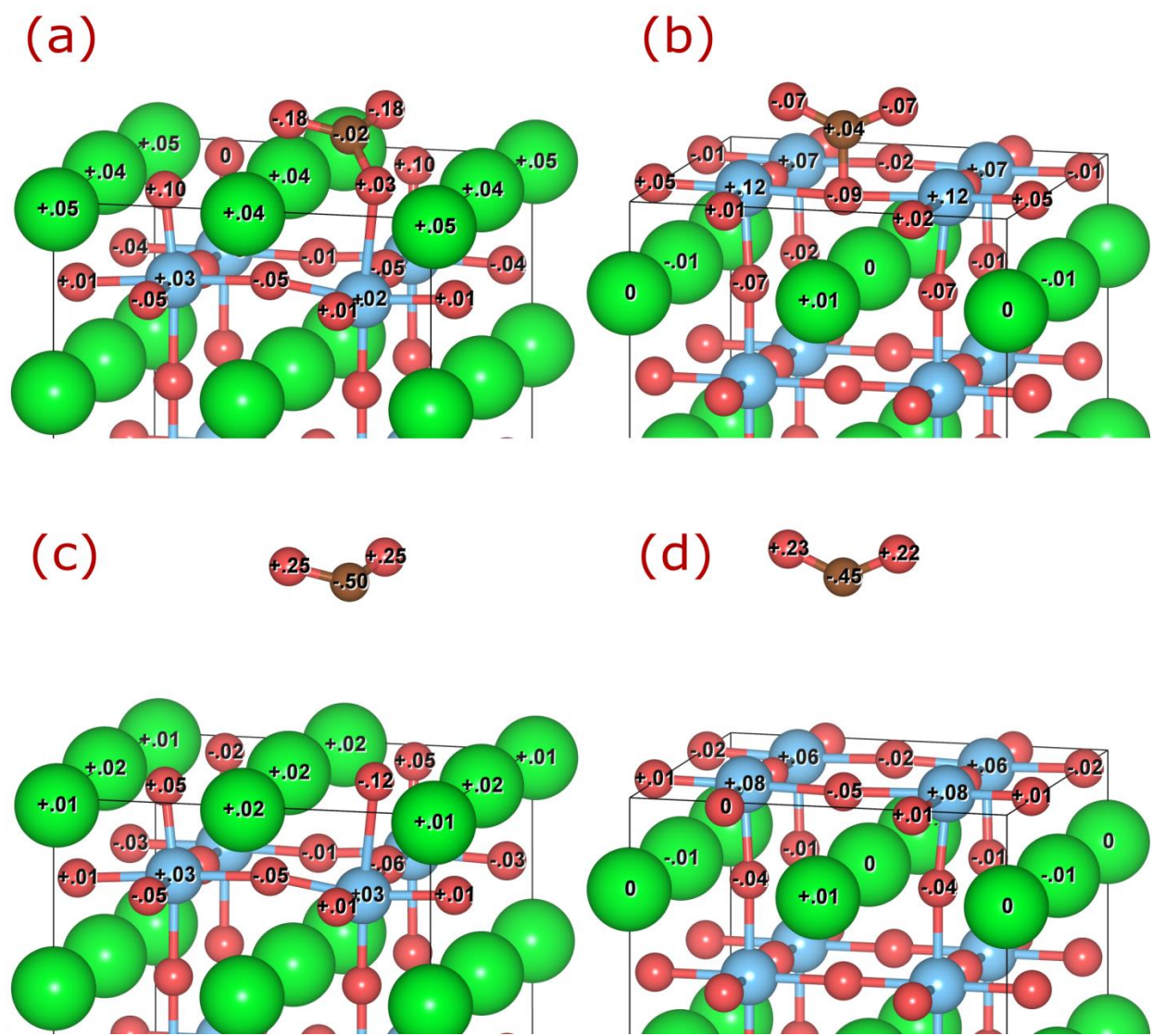

Figure S8. Bader charge transfers imposed by $CO_2$ adsorption on (a) SrO-terminated and (b) $TiO_2$-terminated SrTiO$_3$(001) surfaces. Bader charge redistribution within the separated as-deformed $CO_2$ molecules and (c) SrO-terminated as well as (d) $TiO_2$-terminated SrTiO$_3$(001) surfaces with respect to the free $CO_2$ molecule and clean SrTiO$_3$(001) slabs. The Bader charge transfers of as-deformed $CO_2$ molecules and SrTiO$_3$(001) slabs are computed separately.

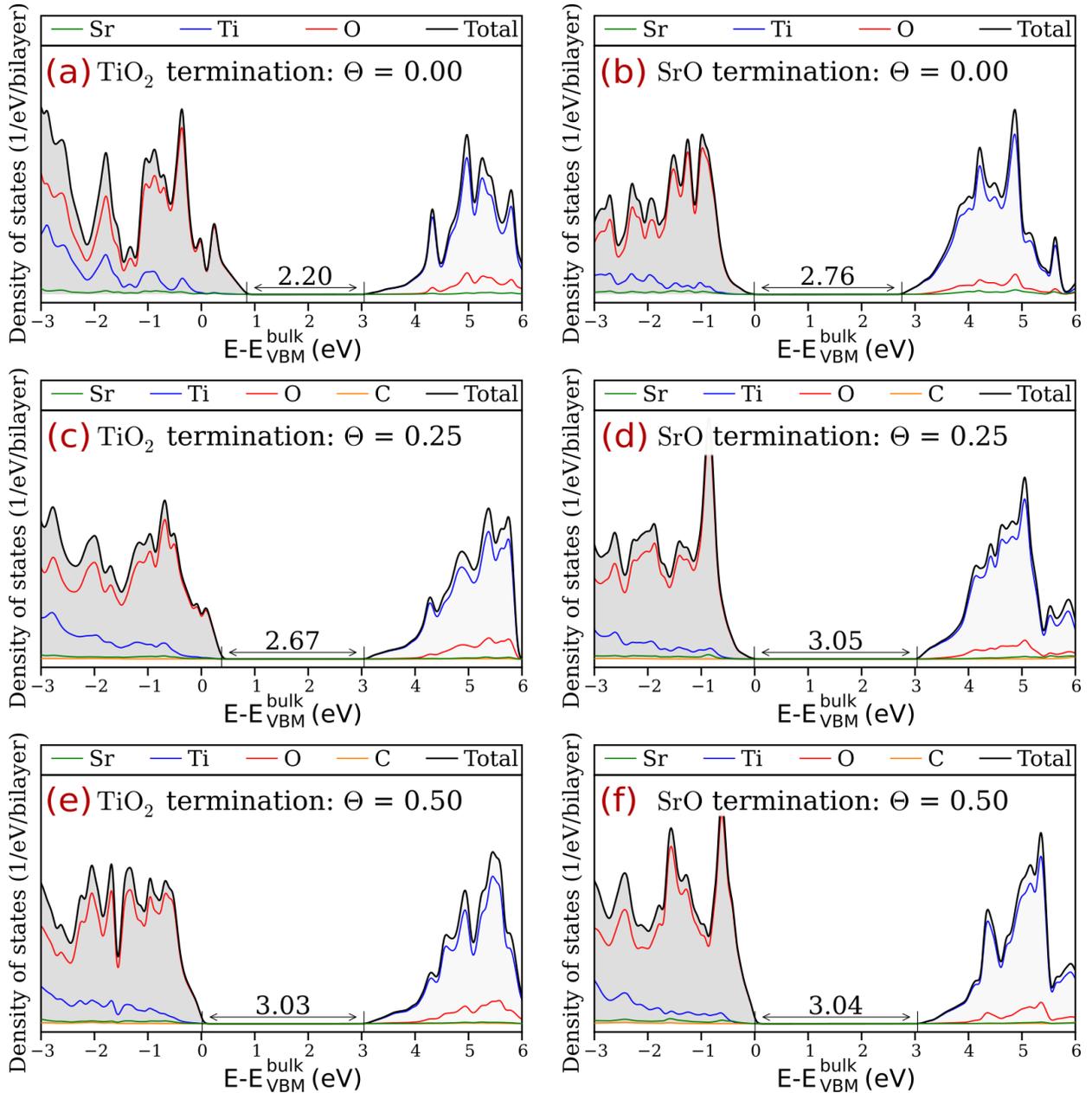

Figure S9. Projected densities of states computed using hybrid HSE functional for two top layers of (a,c,e) TiO$_2$-terminated and (b,d,f) SrO-terminated SrTiO$_3$(001) slabs containing (a,b) zero ($\Theta$=0.00), (c,d) one ($\Theta$=0.25), and (e,f) two ($\Theta$=0.50) adsorbed CO$_2$ molecules per 2×2 surface supercell. For all calculations, we employ 2×2×1 Monkhorst Pack grid and cutoff energy of 400 eV for plane wave basis sets with 3d$^2$4s$^2$ electrons of titanium treated explicitly. The calculations are carried out on the optimized geometries obtained from the PBE relaxations.

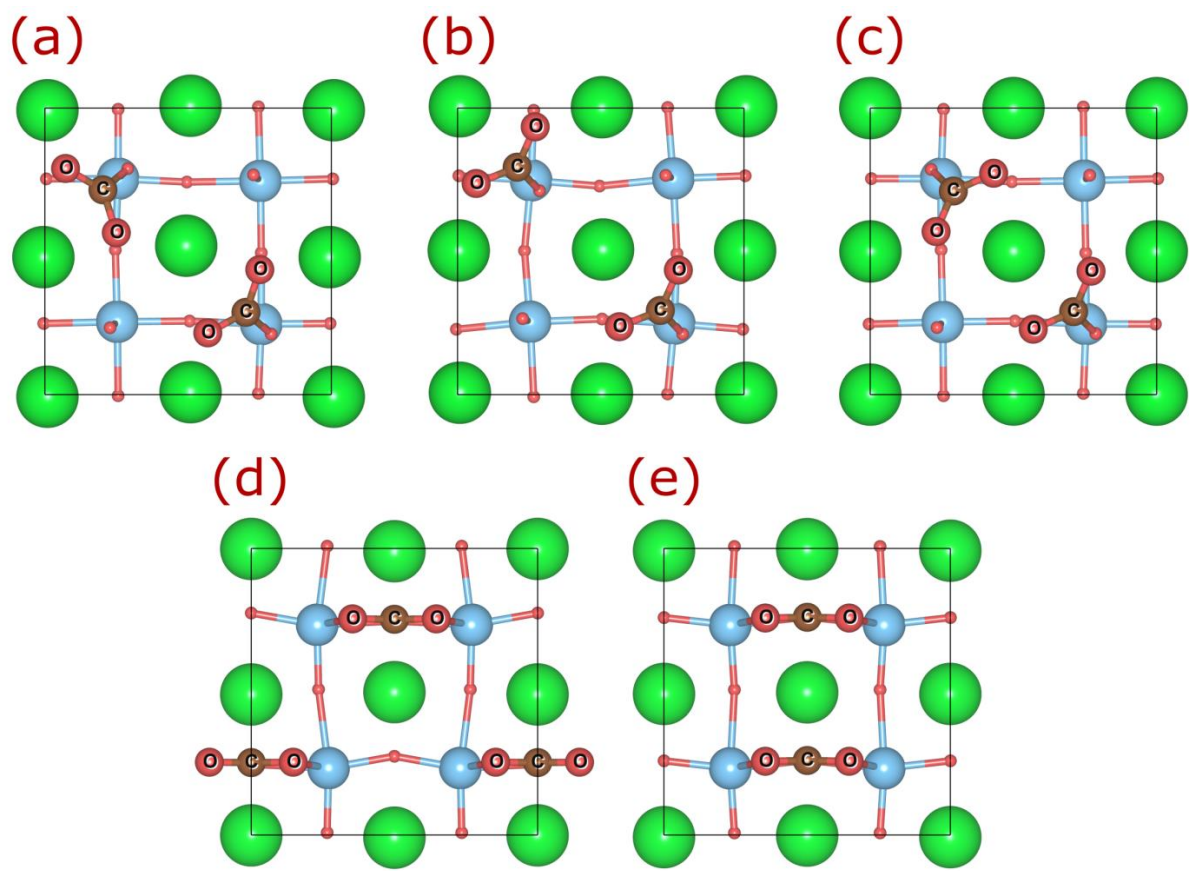

Figure S10. Different $CO_2$ coverage modes corresponding to $\Theta = 0.5$ on (a-c) SrO-terminated and (d,e) TiO$_2$-terminated SrTiO$_3$(001). (a) and (d) are the most stable modes. The computed adsorption energies for (b), (c), and (e) modes are 0.07, 0.14, and 0.03 eV/molecule higher as compared to the most stable configuration.

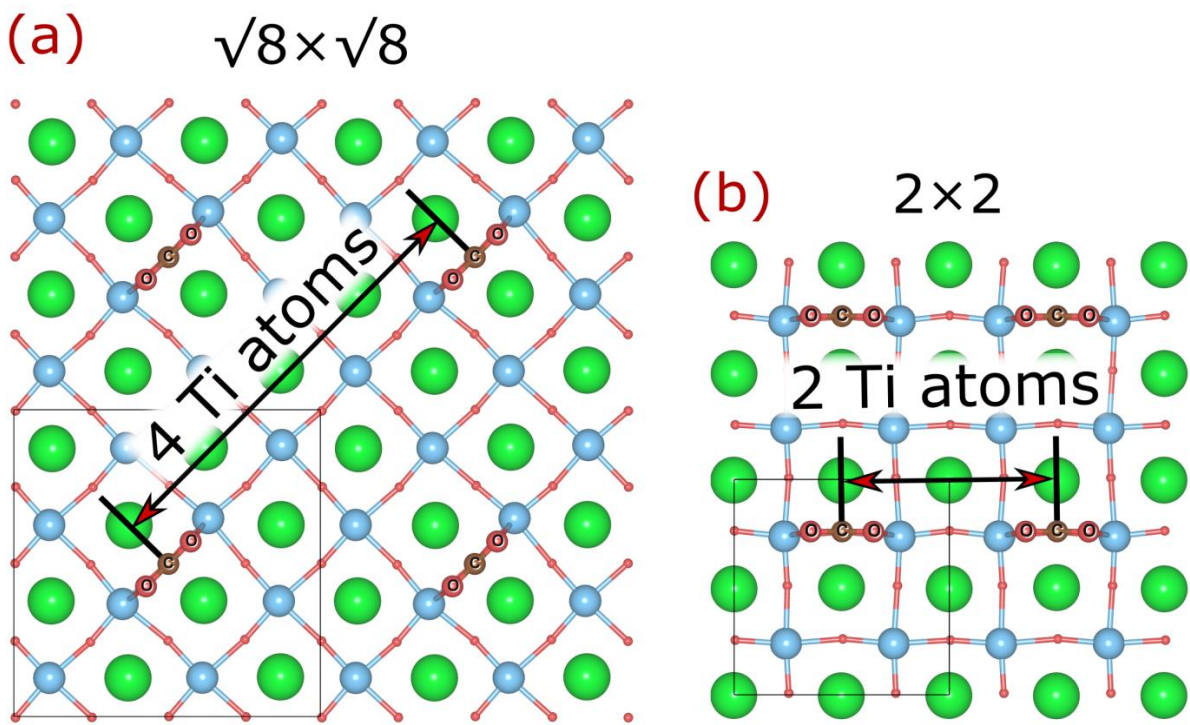

Figure S11. Illustration of "supercell size along O1-O2" term used in this manuscript. O1-O2 refers to a vector connecting O atoms of the adsorbed $CO_2$ molecule. (a) and (b) illustrate the supercell size along O1-O2 for √8×√8 and 2×2 supercells, respectively.

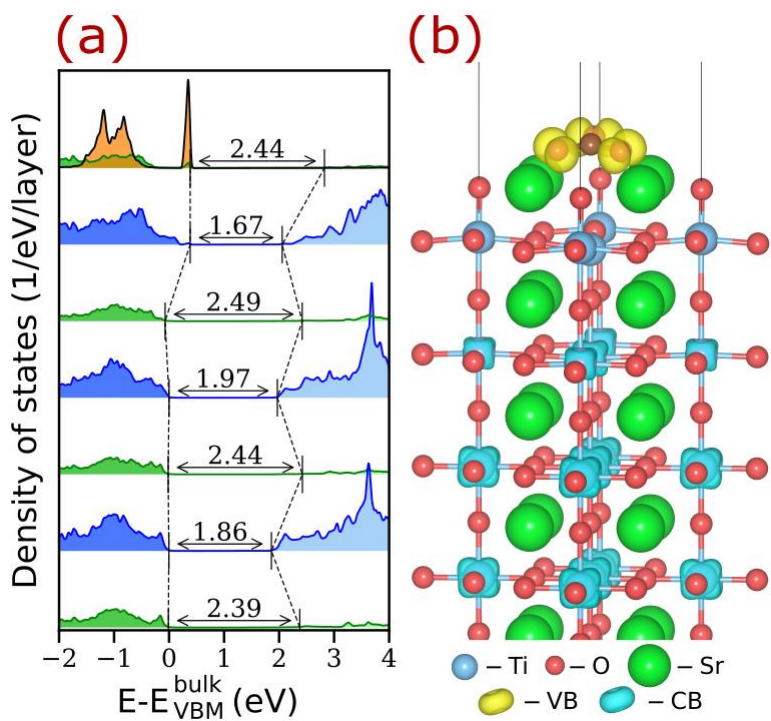

Figure S12. Computed electronic structures of $CO_3$-like complex adsorbed on the surface oxygen vacancy in $\sqrt{2} \times \sqrt{2}$ SrO-terminated $SrTiO_3(001)$ system. (a) Layer-resolved density of states (LDOS). Orange, green, and blue curves represent population densities for the adsorbed $CO_3$-like complex, SrO, and $TiO_2$ layers, respectively. Numbers illustrate effective band gap energies of each atomic layer computed from LDOS neglecting the population densities below 0.1 1/eV/layer. (b) Partial charge density distribution at VB (yellow) and CB (blue) of the $SrTiO_3$ slabs. The computed $CO_2$ adsorption energy for this system is 0.31 eV/molecule higher as compared to the most stable configuration for the same coverage ($\Theta = 0.5$).

Table S1. Sizes of supercells, $\Gamma$-centered Monkhorst-Pack grids used in the calculations, corresponding $CO_2$ coverage, and supercell sizes along O1-O2 atoms of the $CO_2$ molecule adsorbed on $TiO_2$-terminated $SrTiO_3(001)$ surface; see Fig. S11 for more details. The systems are labeled by their lateral dimensions in units of lattice constant. "$2\times2/2$" refers to the systems containing two $CO_2$ molecules per $2\times2$ surface supercell (see Fig. S10).

| Supercell size | Monkhorst-Pack grid | $CO_2$ coverage | Supercell size along O1-O2 |
|---|---|---|---|
| $\sqrt{2}\times\sqrt{2}$ | $7\times7\times1$ | 0.5 | – |
| $1\times2$ | $5\times10\times1$ | 0.5 | 2 |
| $1\times3$ | $3\times10\times1$ | 0.333 | 3 |
| $1\times4$ | $2\times10\times1$ | 0.25 | 4 |
| $1\times5$ | $1\times10\times1$ | 0.2 | 5 |
| $2\times2/2$ | $5\times5\times1$ | 0.5 | 2 |
| $2\times2$ | $5\times5\times1$ | 0.25 | 2 |
| $2\times3$ | $3\times5\times1$ | 0.167 | 3 |
| $2\times4$ | $2\times5\times1$ | 0.125 | 4 |
| $\sqrt{8}\times\sqrt{8}$ | $3\times3\times1$ | 0.125 | 4 |
| $3\times3$ | $3\times3\times1$ | 0.111 | 3 |

Table S2. Structure parameters for the clean $SrTiO_3(001)$ surfaces. The values are given in percent of the lattice constant.

| | SrO-terminated $SrTiO_3(001)$ | $TiO_2$-terminated $SrTiO_3(001)$ |
|---|---|---|
| $S_1$, % | 5.63 | 2.45 |
| $S_2$, % | -1.13 | -3.35 |
| $d_{12}$, % | -6.87 | -6.23 |
| $d_{23}$, % | 3.11 | 4.16 |

Table S3. Structure parameters for the SrTiO$_3$(001) surfaces containing one adsorbed CO$_2$ molecule per 2×2 surface supercell. The values are given in percent of the lattice constant.

| | CO$_2$ on SrO-terminated SrTiO$_3$(001) | CO$_2$ on TiO$_2$-terminated SrTiO$_3$(001) |
|---|---|---|
| $S_1$, % | 4.33 | 0.47 |
| $S_2$, % | -0.01 | -1.59 |
| $d_{12}$, % | -2.89 | -3.39 |
| $d_{23}$, % | 2.01 | 2.05 |

Table S4. Summary of the computational results. The stability (last) column indicates whether the system has the lowest $CO_2$ adsorption energy for a given $CO_2$ coverage.

| Surface | Coverage | System | Figure | | Adsorption energy (eV) | Band gap (eV) | C-Os bond length (Å) | C-O1 (C-O2) bond lengths (Å) | ∠(O1-C-O2) (°) | Stable Yes/No |
|---|---|---|---|---|---|---|---|---|---|---|
| SrO-terminated SrTiO₃(001) | 0.5 | √2×√2 | 2a | - | -1.39 | 1.82 | 1.36 | 1.27 | 123.9 | No |
| | | √2×√2 | S7a | - | -1.15 | 1.43 | 1.30 | 1.30 (1.29) | 119.3 | No |
| | | 2×2/2 | 2a | S10a | -1.46 | 1.84 | 1.36 | 1.27 | 124.3 | Yes |
| | | 2×2/2 | 2a | S10b | -1.39 | 1.84 | 1.36 | 1.27 | 124.0 | No |
| | | 2×2/2 | 2a | S10c | -1.32 | 1.79 | 1.36 | 1.28 | 124.1 | No |
| | 0.25 | 2×2 | 2a | - | -1.94 | 1.82 | 1.33 | 1.29 | 122.2 | Yes |
| | | 2×2 | S7a | - | -1.88 | 1.83 | 1.30 | 1.30 | 119.0 | No |
| | | 2×2 | S7b | - | -1.47 | 1.75 | 1.38 | 1.30 (1.25) | 128.8 | No |
| | | 2×2 | S7c | - | -1.45 | 1.72 | 1.39 | 1.27 | 129.3 | No |
| | | 2×2 | S7d | - | -1.40 | 1.72 | 1.37 | 1.27 | 130.4 | No |
| | 0.167 | 2×3 | S7a | | -2.10 | 1.81 | 1.30 | 1.30 | 119.0 | Yes |
| | | 3×2 | | | -2.09 | 1.82 | 1.30 | 1.30 | 119.0 | No |
| | 0.125 | √8×√8 | | | -2.25 | 1.74 | 1.30 | 1.30 | 119.0 | Yes |
| | 0.111 | 3×3 | | | -2.28 | 1.74 | 1.30 | 1.30 | 119.0 | Yes |
| | 0 | 2×2 | 1f | | - | 1.55 | - | - | - | No |
| TiO₂-terminated SrTiO₃(001) | 0.5 | 1×2 | | - | -1.16 | 1.80 | 1.37 | 1.27 | 131.4 | No |
| | | 2×2/2 | | S10d | -1.19 | 1.85 | 1.37 | 1.27 | 130.8 | Yes |
| | | 2×2/2 | | S10e | -1.16 | 1.80 | 1.37 | 1.27 | 131.4 | No |
| | 0.333 | 1×3 | 2b | - | -1.42 | 1.61 | 1.36 | 1.27 | 130.0 | Yes |
| | 0.25 | 1×4 | | | -1.52 | 1.37 | 1.36 | 1.27 | 129.6 | Yes |
| | | 2×2 | | | -1.24 | 1.50 | 1.37 | 1.27 | 130.8 | No |
| | 0.2 | 1×5 | | | -1.56 | 1.24 | 1.35 | 1.27 | 129.4 | Yes |
| | 0.167 | 2×3 | | | -1.47 | 1.32 | 1.35 | 1.27 | 129.5 | Yes |
| | 0.125 | 2×4 | | | -1.55 | 1.19 | 1.35 | 1.27 | 129.2 | No |
| | | √8×√8 | | | -1.57 | 1.21 | 1.35 | 1.27 | 129.1 | Yes |
| | 0.111 | 3×3 | | | -1.49 | 1.22 | 1.35 | 1.27 | 129.5 | Yes |
| | 0 | 2×2 | 1b | | - | 0.99 | - | - | - | - |

Most stable CO$_2$ adsorption configuration on TiO$_2$-terminated SrTiO$_3$(001) surface as illustrated in Fig. 2b.

```
_chemical_name_common              'Sr24 Ti28 C2 O84 '
_cell_length_a                 7.88820
_cell_length_b                 7.88820
_cell_length_c                 38.64670
_cell_angle_alpha                 90
_cell_angle_beta                  90
_cell_angle_gamma                 90
_space_group_name_H-M_alt        'P 1'

loop_
_space_group_symop_operation_xyz
 'x, y, z'

loop_
  _atom_site_type_symbol
  _atom_site_fract_x
  _atom_site_fract_y
  _atom_site_fract_z
 Sr 0.996244    0.996000    0.049046
 Sr 0.997719    0.487818    0.050107
 Sr 0.496407    0.994011    0.048686
 Sr 0.496961    0.494732    0.048335
 Sr 0.000000    0.000000    0.357185
 Sr 0.000000    0.500000    0.357185
 Sr 0.500000    0.000000    0.357185
 Sr 0.500000    0.500000    0.357185
 Sr 0.497341    0.499463    0.459599
 Sr 0.497436    0.993864    0.459526
 Sr 0.997303    0.499078    0.459529
 Sr 0.997158    0.994234    0.459571
 Sr 0.496961    0.494732    0.563983
 Sr 0.496407    0.994011    0.563632
 Sr 0.997719    0.487818    0.562210
 Sr 0.996244    0.996000    0.563272
 Sr 0.500000    0.500000    0.255129
 Sr 0.500000    0.000000    0.255129
 Sr 0.000000    0.500000    0.255129
 Sr 0.000000    0.000000    0.255129
 Sr 0.997158    0.994234    0.152746
 Sr 0.997303    0.499078    0.152788
 Sr 0.497436    0.993864    0.152791
 Sr 0.497341    0.499463    0.152718
 Ti 0.241796    0.237608    0.003417
```

| | | | |
|---|---|---|---|
| Ti | 0.232222 | 0.736087 | 0.999570 |
| Ti | 0.741241 | 0.237977 | 0.003373 |
| Ti | 0.770702 | 0.735054 | 0.999437 |
| Ti | 0.250000 | 0.250000 | 0.306157 |
| Ti | 0.250000 | 0.750000 | 0.306157 |
| Ti | 0.750000 | 0.250000 | 0.306157 |
| Ti | 0.750000 | 0.750000 | 0.306157 |
| Ti | 0.250000 | 0.250000 | 0.408213 |
| Ti | 0.250000 | 0.750000 | 0.408213 |
| Ti | 0.750000 | 0.250000 | 0.408213 |
| Ti | 0.750000 | 0.750000 | 0.408213 |
| Ti | 0.744185 | 0.742467 | 0.511106 |
| Ti | 0.742844 | 0.242545 | 0.509276 |
| Ti | 0.244124 | 0.742513 | 0.511027 |
| Ti | 0.244457 | 0.242285 | 0.509267 |
| Ti | 0.770702 | 0.735054 | 0.612880 |
| Ti | 0.741241 | 0.237977 | 0.608945 |
| Ti | 0.232222 | 0.736087 | 0.612747 |
| Ti | 0.241796 | 0.237608 | 0.608900 |
| Ti | 0.750000 | 0.750000 | 0.204101 |
| Ti | 0.750000 | 0.250000 | 0.204101 |
| Ti | 0.250000 | 0.750000 | 0.204101 |
| Ti | 0.250000 | 0.250000 | 0.204101 |
| Ti | 0.244457 | 0.242285 | 0.103051 |
| Ti | 0.244124 | 0.742513 | 0.101291 |
| Ti | 0.742844 | 0.242545 | 0.103042 |
| Ti | 0.744185 | 0.742467 | 0.101212 |
| C | 0.501658 | 0.759525 | 0.965227 |
| C | 0.501658 | 0.759525 | 0.647090 |
| O | 0.254871 | 0.007329 | 0.002707 |
| O | 0.245860 | 0.507539 | 0.001269 |
| O | 0.750285 | 0.007641 | 0.002937 |
| O | 0.760750 | 0.506768 | 0.000873 |
| O | 0.005039 | 0.252632 | 0.001663 |
| O | 0.001677 | 0.763749 | 0.997473 |
| O | 0.504944 | 0.260278 | 0.000199 |
| O | 0.502108 | 0.754569 | 0.000566 |
| O | 0.250000 | 0.000000 | 0.306157 |
| O | 0.250000 | 0.500000 | 0.306157 |
| O | 0.750000 | 0.000000 | 0.306157 |
| O | 0.750000 | 0.500000 | 0.306157 |
| O | 0.255807 | 0.256240 | 0.051441 |
| O | 0.250333 | 0.751737 | 0.049890 |
| O | 0.748558 | 0.257714 | 0.051441 |
| O | 0.753866 | 0.749864 | 0.049848 |
| O | 0.000000 | 0.250000 | 0.306157 |

```
O  0.000000    0.750000    0.306157
O  0.500000    0.250000    0.306157
O  0.500000    0.750000    0.306157
O  0.250000    0.000000    0.408213
O  0.250000    0.500000    0.408213
O  0.750000    0.000000    0.408213
O  0.750000    0.500000    0.408213
O  0.250000    0.250000    0.357185
O  0.250000    0.750000    0.357185
O  0.750000    0.250000    0.357185
O  0.750000    0.750000    0.357185
O  0.000000    0.250000    0.408213
O  0.000000    0.750000    0.408213
O  0.500000    0.250000    0.408213
O  0.500000    0.750000    0.408213
O  0.750449    0.752479    0.459490
O  0.752886    0.250720    0.459214
O  0.252213    0.751875    0.459480
O  0.250251    0.251457    0.459215
O  0.502993    0.754427    0.511192
O  0.502699    0.252800    0.509516
O  0.002977    0.752986    0.510546
O  0.003071    0.254171    0.510629
O  0.753866    0.749864    0.562470
O  0.748558    0.257714    0.560876
O  0.250333    0.751737    0.562427
O  0.255807    0.256240    0.560877
O  0.752730    0.503180    0.509811
O  0.753942    0.003351    0.510614
O  0.253915    0.503321    0.510090
O  0.252642    0.003288    0.510354
O  0.502108    0.754569    0.611751
O  0.504944    0.260278    0.612119
O  0.001677    0.763749    0.614844
O  0.005039    0.252632    0.610654
O  0.760750    0.506768    0.611444
O  0.750285    0.007641    0.609380
O  0.245860    0.507539    0.611049
O  0.254871    0.007329    0.609611
O  0.355419    0.759905    0.951604
O  0.647831    0.759728    0.951526
O  0.647831    0.759728    0.660791
O  0.355419    0.759905    0.660713
O  0.750000    0.750000    0.255129
O  0.750000    0.250000    0.255129
O  0.250000    0.750000    0.255129
```

| | | | |
|---|---|---|---|
| O | 0.250000 | 0.250000 | 0.255129 |
| O | 0.500000 | 0.750000 | 0.204101 |
| O | 0.500000 | 0.250000 | 0.204101 |
| O | 0.000000 | 0.750000 | 0.204101 |
| O | 0.000000 | 0.250000 | 0.204101 |
| O | 0.750000 | 0.500000 | 0.204101 |
| O | 0.750000 | 0.000000 | 0.204101 |
| O | 0.250000 | 0.500000 | 0.204101 |
| O | 0.250000 | 0.000000 | 0.204101 |
| O | 0.252642 | 0.003288 | 0.101964 |
| O | 0.253915 | 0.503321 | 0.102228 |
| O | 0.753942 | 0.003351 | 0.101704 |
| O | 0.752730 | 0.503180 | 0.102507 |
| O | 0.003071 | 0.254171 | 0.101689 |
| O | 0.002977 | 0.752986 | 0.101772 |
| O | 0.502699 | 0.252800 | 0.102802 |
| O | 0.502993 | 0.754427 | 0.101126 |
| O | 0.250251 | 0.251457 | 0.153102 |
| O | 0.252213 | 0.751875 | 0.152837 |
| O | 0.752886 | 0.250720 | 0.153103 |
| O | 0.750449 | 0.752479 | 0.152827 |

Most stable $CO_2$ adsorption configuration on SrO-terminated $SrTiO_3(001)$ surface as illustrated in Fig. 2a.

```
_chemical_name_common              'Sr28 Ti24 C2 O80'
_cell_length_a                7.88820
_cell_length_b                7.88820
_cell_length_c                38.64670
_cell_angle_alpha                  90
_cell_angle_beta                   90
_cell_angle_gamma                  90
_space_group_name_H-M_alt          'P 1'

loop_
_space_group_symop_operation_xyz
 'x, y, z'

loop_
  _atom_site_type_symbol
  _atom_site_fract_x
  _atom_site_fract_y
  _atom_site_fract_z
 Sr  0.000000      0.000000      0.4082130
 Sr  0.000000      0.500000      0.4082130
 Sr  0.500000      0.000000      0.4082130
 Sr  0.500000      0.500000      0.4082130
 Sr  0.000000      0.000000      0.3061570
 Sr  0.000000      0.500000      0.3061570
 Sr  0.500000      0.000000      0.3061570
 Sr  0.500000      0.500000      0.3061570
 Sr  0.010784      0.011726      0.0033600
 Sr  0.999082      0.506977      0.0011180
 Sr  0.506349      0.000678      0.0010620
 Sr  0.487067      0.489235      0.9992690
 Sr  0.487067      0.489235      0.6130480
 Sr  0.506349      0.000678      0.6112560
 Sr  0.999082      0.506977      0.6112000
 Sr  0.010784      0.011726      0.6089580
 Sr  0.494203      0.494377      0.5098570
 Sr  0.497871      0.008432      0.5099880
 Sr  0.008262      0.498142      0.5099120
 Sr  0.007404      0.007758      0.5099670
 Sr  0.500000      0.500000      0.2041010
 Sr  0.500000      0.000000      0.2041010
 Sr  0.000000      0.500000      0.2041010
 Sr  0.000000      0.000000      0.2041010
 Sr  0.007404      0.007758      0.1023510
```

| | | | |
|----|----------|----------|----------|
| Sr | 0.008262 | 0.498142 | 0.1024060 |
| Sr | 0.497871 | 0.008432 | 0.1023300 |
| Sr | 0.494203 | 0.494377 | 0.1024610 |
| Ti | 0.250000 | 0.250000 | 0.3571850 |
| Ti | 0.250000 | 0.750000 | 0.3571850 |
| Ti | 0.750000 | 0.250000 | 0.3571850 |
| Ti | 0.750000 | 0.750000 | 0.3571850 |
| Ti | 0.252587 | 0.253451 | 0.0484100 |
| Ti | 0.253884 | 0.751021 | 0.0470410 |
| Ti | 0.750162 | 0.254628 | 0.0470140 |
| Ti | 0.751632 | 0.751993 | 0.0547790 |
| Ti | 0.751632 | 0.751993 | 0.5575380 |
| Ti | 0.750162 | 0.254628 | 0.5653040 |
| Ti | 0.253884 | 0.751021 | 0.5652770 |
| Ti | 0.252587 | 0.253451 | 0.5639070 |
| Ti | 0.752053 | 0.752274 | 0.4577630 |
| Ti | 0.752911 | 0.255830 | 0.4605020 |
| Ti | 0.255645 | 0.753148 | 0.4604970 |
| Ti | 0.251662 | 0.251873 | 0.4593760 |
| Ti | 0.750000 | 0.750000 | 0.2551290 |
| Ti | 0.750000 | 0.250000 | 0.2551290 |
| Ti | 0.250000 | 0.750000 | 0.2551290 |
| Ti | 0.250000 | 0.250000 | 0.2551290 |
| Ti | 0.251662 | 0.251873 | 0.1529410 |
| Ti | 0.255645 | 0.753148 | 0.1518200 |
| Ti | 0.752911 | 0.255830 | 0.1518150 |
| Ti | 0.752053 | 0.752274 | 0.1545540 |
| C | 0.706921 | 0.707332 | 0.966475 |
| C | 0.706921 | 0.707332 | 0.645843 |
| O | 0.250000 | 0.000000 | 0.357185 |
| O | 0.250000 | 0.500000 | 0.357185 |
| O | 0.750000 | 0.000000 | 0.357185 |
| O | 0.750000 | 0.500000 | 0.357185 |
| O | 0.250000 | 0.250000 | 0.408213 |
| O | 0.250000 | 0.750000 | 0.408213 |
| O | 0.750000 | 0.250000 | 0.408213 |
| O | 0.750000 | 0.750000 | 0.408213 |
| O | 0.000000 | 0.250000 | 0.357185 |
| O | 0.000000 | 0.750000 | 0.357185 |
| O | 0.500000 | 0.250000 | 0.357185 |
| O | 0.500000 | 0.750000 | 0.357185 |
| O | 0.249219 | 0.999519 | 0.049875 |
| O | 0.253835 | 0.499771 | 0.051332 |
| O | 0.749698 | 0.994025 | 0.052967 |
| O | 0.746278 | 0.504570 | 0.043052 |
| O | 0.250000 | 0.250000 | 0.306157 |

| O | 0.250000 | 0.750000 | 0.306157 |
|---|----------|----------|----------|
| O | 0.750000 | 0.250000 | 0.306157 |
| O | 0.750000 | 0.750000 | 0.306157 |
| O | 0.999674 | 0.252807 | 0.049863 |
| O | 0.994098 | 0.745570 | 0.052907 |
| O | 0.499940 | 0.250030 | 0.051287 |
| O | 0.504558 | 0.749892 | 0.043175 |
| O | 0.253421 | 0.253325 | 0.999066 |
| O | 0.221377 | 0.745079 | 0.000125 |
| O | 0.746125 | 0.220896 | 0.000124 |
| O | 0.801212 | 0.800763 | 0.987780 |
| O | 0.801212 | 0.800763 | 0.624537 |
| O | 0.746125 | 0.220896 | 0.612193 |
| O | 0.221377 | 0.745079 | 0.612192 |
| O | 0.253421 | 0.253325 | 0.613252 |
| O | 0.504558 | 0.749892 | 0.569142 |
| O | 0.499940 | 0.250030 | 0.561030 |
| O | 0.994098 | 0.745570 | 0.559410 |
| O | 0.999674 | 0.252807 | 0.562454 |
| O | 0.730241 | 0.730134 | 0.511204 |
| O | 0.752202 | 0.263162 | 0.510363 |
| O | 0.263012 | 0.752161 | 0.510347 |
| O | 0.246553 | 0.246361 | 0.510110 |
| O | 0.746278 | 0.504570 | 0.569265 |
| O | 0.749698 | 0.994025 | 0.559350 |
| O | 0.253835 | 0.499771 | 0.560986 |
| O | 0.249219 | 0.999519 | 0.562443 |
| O | 0.499959 | 0.747674 | 0.457984 |
| O | 0.498766 | 0.248441 | 0.459675 |
| O | 0.996133 | 0.749065 | 0.461396 |
| O | 0.998773 | 0.246700 | 0.458908 |
| O | 0.749559 | 0.499869 | 0.457982 |
| O | 0.747621 | 0.996028 | 0.461411 |
| O | 0.247067 | 0.498610 | 0.459678 |
| O | 0.248487 | 0.998661 | 0.458893 |
| O | 0.759086 | 0.557831 | 0.958396 |
| O | 0.557898 | 0.760735 | 0.958385 |
| O | 0.557898 | 0.760735 | 0.653933 |
| O | 0.759086 | 0.557831 | 0.653921 |
| O | 0.750000 | 0.750000 | 0.204101 |
| O | 0.750000 | 0.250000 | 0.204101 |
| O | 0.250000 | 0.750000 | 0.204101 |
| O | 0.250000 | 0.250000 | 0.204101 |
| O | 0.500000 | 0.750000 | 0.255129 |
| O | 0.500000 | 0.250000 | 0.255129 |
| O | 0.000000 | 0.750000 | 0.255129 |

O 0.000000 0.250000 0.255129
O 0.750000 0.500000 0.255129
O 0.750000 0.000000 0.255129
O 0.250000 0.500000 0.255129
O 0.250000 0.000000 0.255129
O 0.248487 0.998661 0.153424
O 0.247067 0.498610 0.152639
O 0.747621 0.996028 0.150906
O 0.749559 0.499869 0.154335
O 0.998773 0.246700 0.153409
O 0.996133 0.749065 0.150921
O 0.498766 0.248441 0.152642
O 0.499959 0.747674 0.154333
O 0.246553 0.246361 0.102208
O 0.263012 0.752161 0.101971
O 0.752202 0.263162 0.101955
O 0.730241 0.730134 0.101114

$CO_2$ adsorption configuration on SrO-terminated $SrTiO_3(001)$ surface as illustrated in Fig. S7a.

```
_chemical_name_common            'Sr28 Ti24 C2 O80'
_cell_length_a              7.88820
_cell_length_b              7.88820
_cell_length_c              38.64670
_cell_angle_alpha               90
_cell_angle_beta                90
_cell_angle_gamma               90
_space_group_name_H-M_alt           'P 1'

loop_
_space_group_symop_operation_xyz
 'x, y, z'

loop_
  _atom_site_type_symbol
  _atom_site_fract_x
  _atom_site_fract_y
  _atom_site_fract_z
 Sr  0.490931     0.506313     0.997930
 Sr  0.490889     0.507466     0.614259
 Sr  0.490012     0.497397     0.509999
 Sr  0.006610     0.507053     0.998033
 Sr  0.006762     0.507421     0.614255
 Sr  0.007344     0.497204     0.510027
 Sr  0.494137     0.009691     0.001783
 Sr  0.493573     0.010400     0.610407
 Sr  0.492829     0.011418     0.510407
 Sr  0.005606     0.010006     0.001902
 Sr  0.005236     0.010698     0.610360
 Sr  0.004296     0.011882     0.510452
 Sr  0.491857     0.496557     0.102241
 Sr  0.009213     0.497255     0.102258
 Sr  0.494907     0.011393     0.101833
 Sr  0.006471     0.011040     0.101887
 Sr  0.499995     0.499995     0.408208
 Sr  0.499995     0.499995     0.306160
 Sr  0.000005     0.499995     0.408208
 Sr  0.000005     0.499995     0.306160
 Sr  0.499995     0.000005     0.408208
 Sr  0.499995     0.000005     0.306160
 Sr  0.000005     0.000005     0.408208
 Sr  0.000005     0.000005     0.306160
 Sr  0.000005     0.000005     0.204104
 Sr  0.499995     0.000005     0.204104
```

| | | | |
|----|----------|----------|----------|
| Sr | 0.000005 | 0.499995 | 0.204104 |
| Sr | 0.499995 | 0.499995 | 0.204104 |
| Ti | 0.249774 | 0.258066 | 0.047710 |
| Ti | 0.249054 | 0.258725 | 0.564512 |
| Ti | 0.248485 | 0.253188 | 0.459463 |
| Ti | 0.749767 | 0.750284 | 0.055515 |
| Ti | 0.749188 | 0.750685 | 0.556733 |
| Ti | 0.746957 | 0.754201 | 0.457451 |
| Ti | 0.250017 | 0.752805 | 0.046882 |
| Ti | 0.749657 | 0.258170 | 0.045623 |
| Ti | 0.748990 | 0.258764 | 0.566575 |
| Ti | 0.248987 | 0.753456 | 0.565377 |
| Ti | 0.748602 | 0.257308 | 0.460801 |
| Ti | 0.246778 | 0.754833 | 0.460519 |
| Ti | 0.250861 | 0.252953 | 0.152830 |
| Ti | 0.752402 | 0.754088 | 0.154853 |
| Ti | 0.252942 | 0.754532 | 0.151783 |
| Ti | 0.750934 | 0.256973 | 0.151476 |
| Ti | 0.250005 | 0.250005 | 0.357184 |
| Ti | 0.750000 | 0.750000 | 0.357184 |
| Ti | 0.250005 | 0.750000 | 0.357184 |
| Ti | 0.750000 | 0.250005 | 0.357184 |
| Ti | 0.750000 | 0.250005 | 0.255129 |
| Ti | 0.250005 | 0.750000 | 0.255129 |
| Ti | 0.750000 | 0.750000 | 0.255129 |
| Ti | 0.250005 | 0.250005 | 0.255129 |
| C  | 0.748240 | 0.703672 | 0.649880 |
| C  | 0.747347 | 0.702997 | 0.962330 |
| O  | 0.249799 | 0.255484 | 0.998622 |
| O  | 0.249770 | 0.255974 | 0.613603 |
| O  | 0.249820 | 0.245844 | 0.510136 |
| O  | 0.500339 | 0.249801 | 0.050329 |
| O  | 0.500516 | 0.250754 | 0.561892 |
| O  | 0.500662 | 0.247824 | 0.459494 |
| O  | 0.998873 | 0.251410 | 0.050246 |
| O  | 0.999099 | 0.251855 | 0.561963 |
| O  | 0.000504 | 0.246672 | 0.459425 |
| O  | 0.250595 | 0.499043 | 0.051792 |
| O  | 0.250366 | 0.499680 | 0.560459 |
| O  | 0.250292 | 0.498157 | 0.459732 |
| O  | 0.248918 | 0.999161 | 0.049383 |
| O  | 0.249229 | 0.999720 | 0.562854 |
| O  | 0.251446 | 0.998486 | 0.458954 |
| O  | 0.750254 | 0.987819 | 0.049656 |
| O  | 0.748572 | 0.506138 | 0.041388 |
| O  | 0.993486 | 0.745613 | 0.047872 |

| O | 0.505496 | 0.747472 | 0.047291 |
|---|----------|----------|----------|
| O | 0.505941 | 0.747866 | 0.564905 |
| O | 0.993781 | 0.746612 | 0.564558 |
| O | 0.749327 | 0.729744 | 0.510749 |
| O | 0.749343 | 0.506656 | 0.570984 |
| O | 0.750550 | 0.988471 | 0.562252 |
| O | 0.502623 | 0.747016 | 0.459524 |
| O | 0.751708 | 0.498936 | 0.458310 |
| O | 0.999273 | 0.748181 | 0.459770 |
| O | 0.750428 | 0.995331 | 0.461201 |
| O | 0.247984 | 0.739921 | 0.999804 |
| O | 0.749487 | 0.218086 | 0.999467 |
| O | 0.749676 | 0.217710 | 0.612705 |
| O | 0.248970 | 0.740552 | 0.612461 |
| O | 0.750449 | 0.258767 | 0.510802 |
| O | 0.251268 | 0.751982 | 0.510352 |
| O | 0.748915 | 0.540073 | 0.653931 |
| O | 0.749008 | 0.539402 | 0.958275 |
| O | 0.891851 | 0.782491 | 0.645925 |
| O | 0.890409 | 0.782564 | 0.966464 |
| O | 0.604037 | 0.781835 | 0.646500 |
| O | 0.602614 | 0.780370 | 0.965545 |
| O | 0.249424 | 0.245525 | 0.102138 |
| O | 0.499270 | 0.247359 | 0.152822 |
| O | 0.999128 | 0.246605 | 0.152849 |
| O | 0.248463 | 0.497905 | 0.152572 |
| O | 0.249200 | 0.998239 | 0.153324 |
| O | 0.749521 | 0.257482 | 0.101462 |
| O | 0.749167 | 0.498714 | 0.153839 |
| O | 0.500472 | 0.746901 | 0.152769 |
| O | 0.747940 | 0.730754 | 0.101524 |
| O | 0.997081 | 0.747675 | 0.152587 |
| O | 0.748401 | 0.995155 | 0.151191 |
| O | 0.249883 | 0.751525 | 0.101945 |
| O | 0.250005 | 0.250005 | 0.408208 |
| O | 0.250005 | 0.250005 | 0.306160 |
| O | 0.499995 | 0.250005 | 0.357184 |
| O | 0.000005 | 0.250005 | 0.357184 |
| O | 0.250005 | 0.499995 | 0.357184 |
| O | 0.250005 | 0.000005 | 0.357184 |
| O | 0.750000 | 0.499995 | 0.357184 |
| O | 0.750000 | 0.750000 | 0.408208 |
| O | 0.499995 | 0.750000 | 0.357184 |
| O | 0.750000 | 0.750000 | 0.306160 |
| O | 0.000005 | 0.750000 | 0.357184 |
| O | 0.750000 | 0.000005 | 0.357184 |

```
O   0.250005   0.750000   0.408208
O   0.750000   0.250005   0.408208
O   0.250005   0.750000   0.306160
O   0.750000   0.250005   0.306160
O   0.750000   0.250005   0.204104
O   0.250005   0.750000   0.204104
O   0.750000   0.000005   0.255129
O   0.000005   0.750000   0.255129
O   0.750000   0.750000   0.204104
O   0.499995   0.750000   0.255129
O   0.750000   0.499995   0.255129
O   0.250005   0.000005   0.255129
O   0.250005   0.499995   0.255129
O   0.000005   0.250005   0.255129
O   0.499995   0.250005   0.255129
O   0.250005   0.250005   0.204104
```

$CO_2$ adsorption configuration on SrO-terminated $SrTiO_3(001)$ surface as illustrated in Fig. S7b.

```
_chemical_name_common            'Sr28 Ti24 C2 O80'
_cell_length_a              7.88820
_cell_length_b              7.88820
_cell_length_c              38.64670
_cell_angle_alpha               90
_cell_angle_beta                90
_cell_angle_gamma               90
_space_group_name_H-M_alt          'P 1'

loop_
_space_group_symop_operation_xyz
 'x, y, z'

loop_
  _atom_site_type_symbol
  _atom_site_fract_x
  _atom_site_fract_y
  _atom_site_fract_z
 Sr  0.000000      0.000000      0.408214
 Sr  0.000000      0.500000      0.408214
 Sr  0.500000      0.000000      0.408214
 Sr  0.500000      0.500000      0.408214
 Sr  0.000000      0.000000      0.306159
 Sr  0.000000      0.500000      0.306159
 Sr  0.500000      0.000000      0.306159
 Sr  0.500000      0.500000      0.306159
 Sr  0.006972      0.997554      0.002133
 Sr  0.006972      0.502446      0.002133
 Sr  0.490068      0.995797      0.001527
 Sr  0.490068      0.504203      0.001527
 Sr  0.490073      0.504185      0.610787
 Sr  0.490073      0.995815      0.610787
 Sr  0.006938      0.502438      0.610174
 Sr  0.006938      0.997562      0.610174
 Sr  0.494897      0.494091      0.509757
 Sr  0.494897      0.005909      0.509757
 Sr  0.000843      0.494153      0.509826
 Sr  0.000843      0.005847      0.509826
 Sr  0.500000      0.500000      0.204103
 Sr  0.500000      0.000000      0.204103
 Sr  0.000000      0.500000      0.204103
 Sr  0.000000      0.000000      0.204103
 Sr  0.000839      0.005866      0.102483
 Sr  0.000839      0.494134      0.102483
```

| | | | |
|---|---|---|---|
| Sr | 0.494898 | 0.005918 | 0.102550 |
| Sr | 0.494898 | 0.494082 | 0.102550 |
| Ti | 0.250000 | 0.250000 | 0.357186 |
| Ti | 0.250000 | 0.750000 | 0.357186 |
| Ti | 0.750000 | 0.250000 | 0.357186 |
| Ti | 0.750000 | 0.750000 | 0.357186 |
| Ti | 0.249319 | 0.250000 | 0.048134 |
| Ti | 0.245003 | 0.750000 | 0.048960 |
| Ti | 0.747771 | 0.250000 | 0.046235 |
| Ti | 0.746656 | 0.750000 | 0.053849 |
| Ti | 0.746650 | 0.750000 | 0.558468 |
| Ti | 0.747734 | 0.250000 | 0.566070 |
| Ti | 0.245010 | 0.750000 | 0.563361 |
| Ti | 0.249306 | 0.250000 | 0.564169 |
| Ti | 0.744854 | 0.750000 | 0.457872 |
| Ti | 0.748369 | 0.250000 | 0.460500 |
| Ti | 0.247283 | 0.750000 | 0.460149 |
| Ti | 0.247367 | 0.250000 | 0.459534 |
| Ti | 0.750000 | 0.750000 | 0.255131 |
| Ti | 0.750000 | 0.250000 | 0.255131 |
| Ti | 0.250000 | 0.750000 | 0.255131 |
| Ti | 0.250000 | 0.250000 | 0.255131 |
| Ti | 0.247373 | 0.250000 | 0.152778 |
| Ti | 0.247250 | 0.750000 | 0.152157 |
| Ti | 0.748377 | 0.250000 | 0.151808 |
| Ti | 0.744822 | 0.750000 | 0.154436 |
| C | 0.742934 | 0.750000 | 0.963407 |
| C | 0.742900 | 0.750000 | 0.648919 |
| O | 0.250000 | 0.000000 | 0.357186 |
| O | 0.250000 | 0.500000 | 0.357186 |
| O | 0.750000 | 0.000000 | 0.357186 |
| O | 0.750000 | 0.500000 | 0.357186 |
| O | 0.250000 | 0.250000 | 0.408214 |
| O | 0.250000 | 0.750000 | 0.408214 |
| O | 0.750000 | 0.250000 | 0.408214 |
| O | 0.750000 | 0.750000 | 0.408214 |
| O | 0.000000 | 0.250000 | 0.357186 |
| O | 0.000000 | 0.750000 | 0.357186 |
| O | 0.500000 | 0.250000 | 0.357186 |
| O | 0.500000 | 0.750000 | 0.357186 |
| O | 0.251646 | 1.000291 | 0.050327 |
| O | 0.251646 | 0.499709 | 0.050327 |
| O | 0.751435 | 0.997204 | 0.048439 |
| O | 0.751435 | 0.502796 | 0.048439 |
| O | 0.250000 | 0.250000 | 0.306159 |
| O | 0.250000 | 0.750000 | 0.306159 |
| O | 0.750000 | 0.250000 | 0.306159 |
| O | 0.750000 | 0.750000 | 0.306159 |
| O | 0.000751 | 0.250000 | 0.049218 |
| O | 0.999350 | 0.750000 | 0.054193 |
| O | 0.501209 | 0.250000 | 0.050733 |
| O | 0.504544 | 0.750000 | 0.045652 |
| O | 0.253929 | 0.250000 | 0.998947 |
| O | 0.229777 | 0.750000 | 0.000842 |
| O | 0.747239 | 0.250000 | 0.998354 |
| O | 0.796138 | 0.750000 | 0.997320 |
| O | 0.579522 | 0.750000 | 0.959437 |
| O | 0.856222 | 0.750000 | 0.940718 |
| O | 0.856199 | 0.750000 | 0.671608 |
| O | 0.579485 | 0.750000 | 0.652889 |
| O | 0.796097 | 0.750000 | 0.615006 |
| O | 0.747206 | 0.250000 | 0.613953 |
| O | 0.229730 | 0.750000 | 0.611474 |
| O | 0.253956 | 0.250000 | 0.613356 |
| O | 0.504518 | 0.750000 | 0.566673 |
| O | 0.501190 | 0.250000 | 0.561572 |
| O | 0.999319 | 0.750000 | 0.558122 |
| O | 0.000734 | 0.250000 | 0.563084 |
| O | 0.734189 | 0.750000 | 0.511165 |
| O | 0.753946 | 0.250000 | 0.510262 |
| O | 0.266193 | 0.750000 | 0.510114 |
| O | 0.247695 | 0.250000 | 0.510251 |
| O | 0.751402 | 0.502793 | 0.563877 |
| O | 0.751402 | 0.997207 | 0.563877 |
| O | 0.251626 | 0.499717 | 0.561980 |
| O | 0.251626 | 1.000283 | 0.561980 |
| O | 0.502672 | 0.750000 | 0.457888 |
| O | 0.500946 | 0.250000 | 0.459669 |
| O | 0.000133 | 0.750000 | 0.461196 |
| O | 0.001017 | 0.250000 | 0.458899 |
| O | 0.751811 | 0.501714 | 0.459697 |
| O | 0.751811 | -0.001714 | 0.459697 |
| O | 0.250922 | 0.500068 | 0.459322 |
| O | 0.250922 | -0.000068 | 0.459322 |
| O | 0.750000 | 0.750000 | 0.204103 |
| O | 0.750000 | 0.250000 | 0.204103 |
| O | 0.250000 | 0.750000 | 0.204103 |
| O | 0.250000 | 0.250000 | 0.204103 |
| O | 0.500000 | 0.750000 | 0.255131 |
| O | 0.500000 | 0.250000 | 0.255131 |
| O | 0.000000 | 0.750000 | 0.255131 |
| O | 0.000000 | 0.250000 | 0.255131 |
| O | 0.750000 | 0.500000 | 0.255131 |
| O | 0.750000 | 0.000000 | 0.255131 |
| O | 0.250000 | 0.500000 | 0.255131 |
| O | 0.250000 | 0.000000 | 0.255131 |
| O | 0.250958 | -0.000070 | 0.152988 |
| O | 0.250958 | 0.500070 | 0.152988 |
| O | 0.751849 | -0.001717 | 0.152615 |
| O | 0.751849 | 0.501717 | 0.152615 |
| O | 0.001038 | 0.250000 | 0.153419 |
| O | 0.000169 | 0.750000 | 0.151103 |
| O | 0.500970 | 0.250000 | 0.152634 |
| O | 0.502689 | 0.750000 | 0.154422 |
| O | 0.247667 | 0.250000 | 0.102059 |
| O | 0.266227 | 0.750000 | 0.102191 |
| O | 0.754011 | 0.250000 | 0.102047 |
| O | 0.734210 | 0.750000 | 0.101152 |

$CO_2$ adsorption configuration on SrO-terminated $SrTiO_3(001)$ surface as illustrated in Fig. S7c.

```
_chemical_name_common              'Sr28 Ti24 C2 O80'
_cell_length_a              7.88820
_cell_length_b              7.88820
_cell_length_c              38.64670
_cell_angle_alpha              90
_cell_angle_beta              90
_cell_angle_gamma              90
_space_group_name_H-M_alt          'P 1'

loop_
_space_group_symop_operation_xyz
  'x, y, z'

loop_
   _atom_site_type_symbol
   _atom_site_fract_x
   _atom_site_fract_y
   _atom_site_fract_z
 Sr  0.000000      0.000000      0.408214
 Sr  0.000000      0.500000      0.408214
 Sr  0.500000      0.000000      0.408214
 Sr  0.500000      0.500000      0.408214
 Sr  0.000000      0.000000      0.306159
 Sr  0.000000      0.500000      0.306159
 Sr  0.500000      0.000000      0.306159
 Sr  0.500000      0.500000      0.306159
 Sr -0.002230     -0.002230      0.000080
 Sr  0.004021      0.496354      0.003951
 Sr  0.496354      0.004021      0.003951
 Sr  0.502575      0.502575      0.000024
 Sr  0.502571      0.502571      0.612283
 Sr  0.496351      0.004026      0.608357
 Sr  0.004026      0.496351      0.608357
 Sr -0.002235     -0.002235      0.612230
 Sr  0.495485      0.495485      0.509980
 Sr  0.495693      0.004483      0.509483
 Sr  0.004483      0.495693      0.509483
 Sr  0.004691      0.004691      0.509979
 Sr  0.500000      0.500000      0.204103
 Sr  0.500000      0.000000      0.204103
 Sr  0.000000      0.500000      0.204103
 Sr  0.000000      0.000000      0.204103
 Sr  0.004699      0.004699      0.102332
 Sr  0.004489      0.495689      0.102828
```

| | | | |
|---|---|---|---|
| Sr | 0.495689 | 0.004489 | 0.102828 |
| Sr | 0.495481 | 0.495481 | 0.102330 |
| Ti | 0.250000 | 0.250000 | 0.357186 |
| Ti | 0.250000 | 0.750000 | 0.357186 |
| Ti | 0.750000 | 0.250000 | 0.357186 |
| Ti | 0.750000 | 0.750000 | 0.357186 |
| Ti | 0.250169 | 0.250169 | 0.048031 |
| Ti | 0.250197 | 0.750134 | 0.047749 |
| Ti | 0.750134 | 0.250197 | 0.047749 |
| Ti | 0.750152 | 0.750152 | 0.054036 |
| Ti | 0.750149 | 0.750149 | 0.558273 |
| Ti | 0.750135 | 0.250192 | 0.564554 |
| Ti | 0.250192 | 0.750135 | 0.564554 |
| Ti | 0.250166 | 0.250166 | 0.564273 |
| Ti | 0.750072 | 0.750072 | 0.457767 |
| Ti | 0.750077 | 0.250069 | 0.460250 |
| Ti | 0.250069 | 0.750077 | 0.460250 |
| Ti | 0.250071 | 0.250071 | 0.459567 |
| Ti | 0.750000 | 0.750000 | 0.255131 |
| Ti | 0.750000 | 0.250000 | 0.255131 |
| Ti | 0.250000 | 0.750000 | 0.255131 |
| Ti | 0.250000 | 0.250000 | 0.255131 |
| Ti | 0.250073 | 0.250073 | 0.152748 |
| Ti | 0.250070 | 0.750078 | 0.152057 |
| Ti | 0.750078 | 0.250070 | 0.152057 |
| Ti | 0.750075 | 0.750075 | 0.154545 |
| C | 0.750101 | 0.750101 | 0.961298 |
| C | 0.750110 | 0.750110 | 0.651018 |
| O | 0.250000 | 0.000000 | 0.357186 |
| O | 0.250000 | 0.500000 | 0.357186 |
| O | 0.750000 | 0.000000 | 0.357186 |
| O | 0.750000 | 0.500000 | 0.357186 |
| O | 0.250000 | 0.250000 | 0.408214 |
| O | 0.250000 | 0.750000 | 0.408214 |
| O | 0.750000 | 0.250000 | 0.408214 |
| O | 0.750000 | 0.750000 | 0.408214 |
| O | 0.000000 | 0.250000 | 0.357186 |
| O | 0.000000 | 0.750000 | 0.357186 |
| O | 0.500000 | 0.250000 | 0.357186 |
| O | 0.500000 | 0.750000 | 0.357186 |
| O | 0.251390 | 1.000047 | 0.050339 |
| O | 0.248624 | 0.499959 | 0.050373 |
| O | 0.749451 | 0.997653 | 0.049305 |
| O | 0.750587 | 0.502395 | 0.049339 |
| O | 0.250000 | 0.250000 | 0.306159 |
| O | 0.250000 | 0.750000 | 0.306159 |
| O | 0.750000 | 0.250000 | 0.306159 |
| O | 0.750000 | 0.750000 | 0.306159 |
| O | 1.000047 | 0.251390 | 0.050339 |
| O | 0.997653 | 0.749451 | 0.049305 |
| O | 0.499959 | 0.248624 | 0.050373 |
| O | 0.502395 | 0.750587 | 0.049339 |
| O | 0.250060 | 0.250060 | 0.999140 |
| O | 0.250099 | 0.749847 | -0.000397 |
| O | 0.749847 | 0.250099 | -0.000397 |
| O | 0.749826 | 0.749826 | 0.997135 |
| O | 0.853192 | 0.853192 | 0.947269 |
| O | 0.647234 | 0.647234 | 0.947149 |
| O | 0.647220 | 0.647220 | 0.665161 |
| O | 0.853183 | 0.853183 | 0.665052 |
| O | 0.749856 | 0.749856 | 0.615180 |
| O | 0.749847 | 0.250084 | 0.612701 |
| O | 0.250084 | 0.749847 | 0.612701 |
| O | 0.250064 | 0.250064 | 0.613164 |
| O | 0.502403 | 0.750590 | 0.562969 |
| O | 0.499961 | 0.248626 | 0.561933 |
| O | 0.997649 | 0.749451 | 0.562996 |
| O | 1.000049 | 0.251393 | 0.561967 |
| O | 0.750087 | 0.750087 | 0.511095 |
| O | 0.750025 | 0.249958 | 0.510161 |
| O | 0.249958 | 0.750025 | 0.510161 |
| O | 0.249952 | 0.249952 | 0.510058 |
| O | 0.750590 | 0.502403 | 0.562969 |
| O | 0.749451 | 0.997649 | 0.562996 |
| O | 0.248626 | 0.499961 | 0.561933 |
| O | 0.251393 | 1.000049 | 0.561967 |
| O | 0.501607 | 0.750022 | 0.459598 |
| O | 0.500097 | 0.249974 | 0.459244 |
| O | -0.001613 | 0.749951 | 0.459584 |
| O | -0.000117 | 0.250018 | 0.459237 |
| O | 0.750022 | 0.501607 | 0.459598 |
| O | 0.749951 | -0.001613 | 0.459584 |
| O | 0.249974 | 0.500097 | 0.459244 |
| O | 0.250018 | -0.000117 | 0.459237 |
| O | 0.750000 | 0.750000 | 0.204103 |
| O | 0.750000 | 0.250000 | 0.204103 |
| O | 0.250000 | 0.750000 | 0.204103 |
| O | 0.250000 | 0.250000 | 0.204103 |
| O | 0.500000 | 0.750000 | 0.255131 |
| O | 0.500000 | 0.250000 | 0.255131 |
| O | 0.000000 | 0.750000 | 0.255131 |
| O | 0.000000 | 0.250000 | 0.255131 |
| O | 0.750000 | 0.500000 | 0.255131 |
| O | 0.750000 | 0.000000 | 0.255131 |
| O | 0.250000 | 0.500000 | 0.255131 |
| O | 0.250000 | 0.000000 | 0.255131 |
| O | 0.250018 | -0.000118 | 0.153077 |
| O | 0.249974 | 0.500097 | 0.153070 |
| O | 0.749949 | -0.001618 | 0.152728 |
| O | 0.750021 | 0.501611 | 0.152712 |
| O | -0.000118 | 0.250018 | 0.153077 |
| O | -0.001618 | 0.749949 | 0.152728 |
| O | 0.500097 | 0.249974 | 0.153070 |
| O | 0.501611 | 0.750021 | 0.152712 |
| O | 0.249952 | 0.249952 | 0.102254 |
| O | 0.249946 | 0.750022 | 0.102150 |
| O | 0.750022 | 0.249946 | 0.102150 |
| O | 0.750100 | 0.750100 | 0.101212 |

CO$_2$ adsorption configuration on SrO-terminated SrTiO$_3$(001) surface as illustrated in Fig. S7d.

```
_chemical_name_common              'Sr28 Ti24 C2 O80'
_cell_length_a              7.88820
_cell_length_b              7.88820
_cell_length_c              38.64670
_cell_angle_alpha              90
_cell_angle_beta              90
_cell_angle_gamma              90
_space_group_name_H-M_alt          'P 1'

loop_
_space_group_symop_operation_xyz
  'x, y, z'

loop_
  _atom_site_type_symbol
  _atom_site_fract_x
  _atom_site_fract_y
  _atom_site_fract_z
 Sr  0.000000      0.000000      0.408214
 Sr  0.000000      0.500000      0.408214
 Sr  0.500000      0.000000      0.408214
 Sr  0.500000      0.500000      0.408214
 Sr  0.000000      0.000000      0.306159
 Sr  0.000000      0.500000      0.306159
 Sr  0.500000      0.000000      0.306159
 Sr  0.500000      0.500000      0.306159
 Sr  0.006816      0.991668      0.001736
 Sr  0.006816      0.508332      0.001736
 Sr  0.493184      0.991668      0.001736
 Sr  0.493184      0.508332      0.001736
 Sr  0.493187      0.508322      0.610572
 Sr  0.493187      0.991678      0.610572
 Sr  0.006813      0.508322      0.610572
 Sr  0.006813      0.991678      0.610572
 Sr  0.496563      0.493597      0.509805
 Sr  0.496563      0.006403      0.509805
 Sr  0.003437      0.493597      0.509805
 Sr  0.003437      0.006403      0.509805
 Sr  0.500000      0.500000      0.204103
 Sr  0.500000      0.000000      0.204103
 Sr  0.000000      0.500000      0.204103
 Sr  0.000000      0.000000      0.204103
 Sr  0.003441      0.006408      0.102504
 Sr  0.003441      0.493592      0.102504
```

| | | |
|---|---|---|
| Sr | 0.496559 | 0.006408 | 0.102504 |
| Sr | 0.496559 | 0.493592 | 0.102504 |
| Ti | 0.250000 | 0.250000 | 0.357186 |
| Ti | 0.250000 | 0.750000 | 0.357186 |
| Ti | 0.750000 | 0.250000 | 0.357186 |
| Ti | 0.750000 | 0.750000 | 0.357186 |
| Ti | 0.250000 | 0.250000 | 0.047397 |
| Ti | 0.250000 | 0.750000 | 0.049380 |
| Ti | 0.750000 | 0.250000 | 0.046033 |
| Ti | 0.750000 | 0.750000 | 0.054419 |
| Ti | 0.750000 | 0.750000 | 0.557891 |
| Ti | 0.750000 | 0.250000 | 0.566273 |
| Ti | 0.250000 | 0.750000 | 0.562928 |
| Ti | 0.250000 | 0.250000 | 0.564903 |
| Ti | 0.750000 | 0.750000 | 0.457715 |
| Ti | 0.750000 | 0.250000 | 0.460522 |
| Ti | 0.250000 | 0.750000 | 0.459889 |
| Ti | 0.250000 | 0.250000 | 0.459678 |
| Ti | 0.750000 | 0.750000 | 0.255131 |
| Ti | 0.750000 | 0.250000 | 0.255131 |
| Ti | 0.250000 | 0.750000 | 0.255131 |
| Ti | 0.250000 | 0.250000 | 0.255131 |
| Ti | 0.250000 | 0.250000 | 0.152633 |
| Ti | 0.250000 | 0.750000 | 0.152421 |
| Ti | 0.750000 | 0.250000 | 0.151786 |
| Ti | 0.750000 | 0.750000 | 0.154595 |
| C | 0.750000 | 0.750000 | 0.963359 |
| C | 0.750000 | 0.750000 | 0.648966 |
| O | 0.250000 | 0.000000 | 0.357186 |
| O | 0.250000 | 0.500000 | 0.357186 |
| O | 0.750000 | 0.000000 | 0.357186 |
| O | 0.750000 | 0.500000 | 0.357186 |
| O | 0.250000 | 0.250000 | 0.408214 |
| O | 0.250000 | 0.750000 | 0.408214 |
| O | 0.750000 | 0.250000 | 0.408214 |
| O | 0.750000 | 0.750000 | 0.408214 |
| O | 0.000000 | 0.250000 | 0.357186 |
| O | 0.000000 | 0.750000 | 0.357186 |
| O | 0.500000 | 0.250000 | 0.357186 |
| O | 0.500000 | 0.750000 | 0.357186 |
| O | 0.250000 | 1.000493 | 0.050155 |
| O | 0.250000 | 0.499507 | 0.050155 |
| O | 0.750000 | 0.997216 | 0.048751 |
| O | 0.750000 | 0.502784 | 0.048751 |
| O | 0.250000 | 0.250000 | 0.306159 |
| O | 0.250000 | 0.750000 | 0.306159 |
| O | 0.750000 | 0.250000 | 0.306159 |
| O | 0.750000 | 0.750000 | 0.306159 |
| O | 0.999728 | 0.250000 | 0.049698 |
| O | 0.998439 | 0.750000 | 0.050296 |
| O | 0.500272 | 0.250000 | 0.049698 |
| O | 0.501561 | 0.750000 | 0.050296 |
| O | 0.250000 | 0.250000 | 0.998607 |
| O | 0.250000 | 0.750000 | 0.000694 |
| O | 0.750000 | 0.250000 | 0.998224 |
| O | 0.750000 | 0.750000 | 0.998823 |
| O | 0.603290 | 0.750000 | 0.949551 |
| O | 0.896710 | 0.750000 | 0.949551 |
| O | 0.896705 | 0.750000 | 0.662776 |
| O | 0.603295 | 0.750000 | 0.662776 |
| O | 0.750000 | 0.750000 | 0.613498 |
| O | 0.750000 | 0.250000 | 0.614084 |
| O | 0.250000 | 0.750000 | 0.611614 |
| O | 0.250000 | 0.250000 | 0.613694 |
| O | 0.501567 | 0.750000 | 0.562011 |
| O | 0.500274 | 0.250000 | 0.562607 |
| O | 0.998433 | 0.750000 | 0.562011 |
| O | 0.999726 | 0.250000 | 0.562607 |
| O | 0.750000 | 0.750000 | 0.510819 |
| O | 0.750000 | 0.250000 | 0.510228 |
| O | 0.250000 | 0.750000 | 0.510141 |
| O | 0.250000 | 0.250000 | 0.510185 |
| O | 0.750000 | 0.502781 | 0.563558 |
| O | 0.750000 | 0.997219 | 0.563558 |
| O | 0.250000 | 0.499512 | 0.562147 |
| O | 0.250000 | 1.000488 | 0.562147 |
| O | 0.501351 | 0.750000 | 0.459480 |
| O | 0.499929 | 0.250000 | 0.459298 |
| O | -0.001351 | 0.750000 | 0.459480 |
| O | 0.000071 | 0.250000 | 0.459298 |
| O | 0.750000 | 0.501653 | 0.459595 |
| O | 0.750000 | -0.001653 | 0.459595 |
| O | 0.250000 | 0.500275 | 0.459292 |
| O | 0.250000 | -0.000275 | 0.459292 |
| O | 0.750000 | 0.750000 | 0.204103 |
| O | 0.750000 | 0.250000 | 0.204103 |
| O | 0.250000 | 0.750000 | 0.204103 |
| O | 0.250000 | 0.250000 | 0.204103 |
| O | 0.500000 | 0.750000 | 0.255131 |
| O | 0.500000 | 0.250000 | 0.255131 |
| O | 0.000000 | 0.750000 | 0.255131 |
| O | 0.000000 | 0.250000 | 0.255131 |

```
O  0.750000      0.500000      0.255131
O  0.750000      0.000000      0.255131
O  0.250000      0.500000      0.255131
O  0.250000      0.000000      0.255131
O  0.250000     -0.000277      0.153020
O  0.250000      0.500277      0.153020
O  0.750000     -0.001656      0.152715
O  0.750000      0.501656      0.152715
O  0.000070      0.250000      0.153015
O -0.001352      0.750000      0.152830
O  0.499930      0.250000      0.153015
O  0.501352      0.750000      0.152830
O  0.250000      0.250000      0.102125
O  0.250000      0.750000      0.102167
O  0.750000      0.250000      0.102082
O  0.750000      0.750000      0.101490
```